\documentclass[lettersize,journal]{IEEEtran}
\usepackage{amsmath,graphicx}
\usepackage[linesnumbered,ruled]{algorithm2e}
\usepackage{multirow}
\usepackage{hhline}
\usepackage{cite}
\usepackage{pifont}
\usepackage{bm}
\usepackage{xcolor}
\usepackage{amsmath}

\newcommand{\tim}[1]{{\textcolor{blue}{#1}}}

\usepackage[pagebackref=false,colorlinks=true,bookmarks=false,linkcolor=red,citecolor=blue,urlcolor=red]{hyperref}

\usepackage{tabularx} 
\newcolumntype{Y}{>{\centering\arraybackslash}X}

\ifCLASSINFOpdf
\else
\fi
\begin{document}
\bstctlcite{IEEEexample:BSTcontrol}
%
\title{Exploring Cross-Utterance Speech Contexts for Conformer-Transducer Speech Recognition Systems}
%
%

%

%
%
\author{Mingyu Cui, Mengzhe Geng, Jiajun Deng, Chengxi Deng, Jiawen Kang, Shujie Hu, Guinan Li, Tianzi Wang, Zhaoqing Li, Xie Chen,~\IEEEmembership{Member,~IEEE}, Xunying Liu,~\IEEEmembership{Member,~IEEE}
\thanks{Mingyu Cui, Jiajun Deng, Chengxi Deng, Jiawen Kang, Shujie Hu, Guinan Li, Tianzi Wang, Zhaoqing Li are with the Chinese University of Hong Kong, China (email: \{mycui, jjdeng, cxdeng, jwkang, sjhu, gnli, twang, zqli\}@se.cuhk.edu.hk)}
\thanks{Mengzhe Geng is with the National Research Council Canada, Canada (email: Mengzhe.Geng@nrc-cnrc.gc.ca);}
\thanks{Xie Chen is with the Shanghai Jiao Tong University, China (email: chenxie95@sjtu.edu.cn)}
\thanks{Xunying Liu is with the Chinese University of Hong Kong, China and the corresponding author (email:  xyliu@se.cuhk.edu.hk).}}

\markboth{IEEE Transactions on Audio, Speech and Language Processing}%
{Bare Demo of IEEEtran.cls for IEEE Journals}

%



\maketitle

\begin{abstract}

This paper investigates four types of cross-utterance speech contexts modeling approaches for streaming and non-streaming Conformer-Transformer (C-T) ASR systems: i) input audio feature concatenation; ii) cross-utterance Encoder embedding concatenation; iii) cross-utterance Encoder embedding pooling projection; or iv) a novel chunk-based approach applied to C-T models for the first time. An efficient batch-training scheme is proposed for contextual C-Ts that uses spliced speech utterances within each minibatch to minimize the synchronization overhead while preserving the sequential order of cross-utterance speech contexts. Experiments are conducted on four benchmark speech datasets across three languages: the English GigaSpeech and Mandarin Wenetspeech corpora used in contextual C-T models pre-training; and the English DementiaBank Pitt and Cantonese JCCOCC MoCA elderly speech datasets used in domain fine-tuning. The best performing contextual C-T systems consistently outperform their respective baselines using no cross-utterance speech contexts in pre-training and fine-tuning stages with statistically significant average word error rate (WER) or character error rate (CER) reductions up to \textbf{0.9\%}, \textbf{1.1\%}, \textbf{0.51\%}, and \textbf{0.98\%} absolute (\textbf{6.0\%}, \textbf{5.4\%}, \textbf{2.0\%}, and \textbf{3.4\%} relative) on the four tasks respectively. Their performance competitiveness against Wav2vec2.0-Conformer, XLSR-128, and Whisper models highlights the potential benefit of incorporating cross-utterance speech contexts into current speech foundation models. 

\end{abstract}


%
\IEEEpeerreviewmaketitle
\vspace{-0.3cm}
\section{Introduction}
\label{introduction}
\IEEEPARstart{E}{nd-to-end (E2E)} automatic speech recognition (ASR) technologies have achieved great success in recent years~\cite{graves2006connectionist, watanabe2017hybrid,chan2016listen,vaswani2017attention,dong2018speech, karita2019comparative, gulati2020conformer,guo2021recent,graves2012sequence, medsker2001recurrent, sak2014long, zhou2022efficient}.
Transformer-based ASR models, in particular, those based on Conformer Encoder architectures \cite{tuske2021limit,zeineldeen2022improving, deng2022confidence, saon2023diagonal, li2022recent}, have produced superior
performance across various ASR tasks.
\vspace{-0.4cm} 
\subsection{Cross-utterance Speech Contexts Modeling for ASR}
\label{intro_approach}
Context plays an important role in human communication. Rich contextual cues across neighbouring speech utterances at acoustic-phonetic, prosodic, lexical and semantic level are used to determine what is said in a natural conversation. However, the majority of current ASR systems are trained and evaluated at the single utterance level,  while longer range cross-utterance speech contexts are not fully utilized. 

\begin{figure}
    \centering
\includegraphics[width=0.9\linewidth]{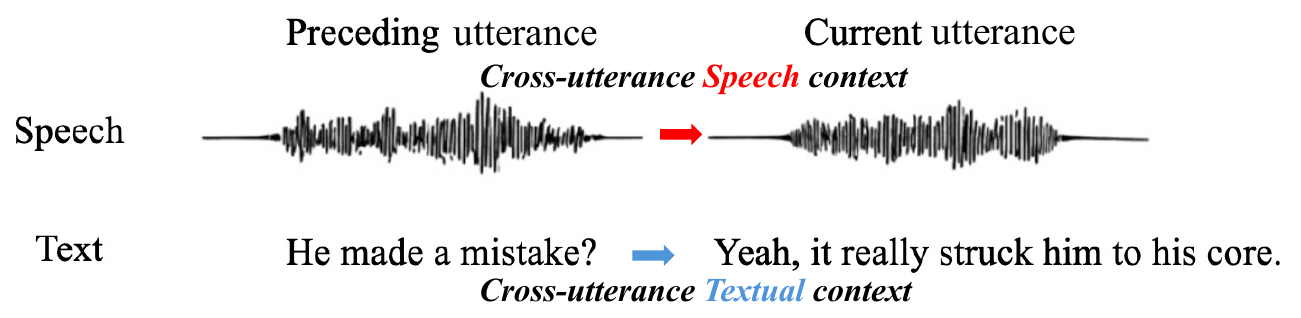}
    \caption{\textcolor{black}{An example of cross-utterance speech contexts (red color) and cross-utterance textual contexts (blue color).}}
    \label{fig:example}
\end{figure}

To this end, existing researches have largely focused on incorporating two
broad types of cross-utterance   contexts into these systems: \textit{\textbf{1) cross-utterance textual contexts}}, which have been widely utilized in the Decoder or Predictor modules of E2E ASR systems such as attention based Encoder-Decoder (AED) \cite{wei2022improving, tang2024improvingasrcontextualbiasing, hsiao2024optimizing, jiang2024contextual, yang2024promptasr, wei2024conversational, wu2023dual, zhang2023cppf, shi2023casa, chan2023using} models and neural Transducer systems \cite{hou2022bring, zhou2023copyne,huang2024improving}, as well as separately constructed language models \cite{irie2019training, chen2020lstm, xiong2018session, dai2019transformer, kim2019end, liu2020contextualizing, liu2013use, beltagy2020longformer, sun2021transformer}, or large language model (LLM) based ASR systems \cite{yang2024ctc, bai2024seed, asano2025contextual, everson2024towards, lei2025contextualization, lakomkin2024end, suhimproving, yang2024mala, tang2024contextualized, manh2024improving, li2024using}; and \textit{\textbf{2) cross-utterance speech contexts}}, which are more directly related to the audio and often integrated into the Encoder components of E2E systems \cite{rae2019compressive, fan2019speaker, tsunoo2019transformer, hori2020transformer, hori2021advanced}. \textcolor{black}{An example of cross-utterance speech contexts and cross-utterance textual contexts is shown in Figure \ref{fig:example}.}

Among the above two, the importance of \textit{\textbf{cross-utterance speech contexts}} has been relatively underappreciated. Prior researches in this direction can be characterized using two key aspects: \textbf{2a) Backbone E2E architectures} that are based on either RNNs \cite{chang2023context, gong2023longfnt, gong2024advanced, kim2019gated, kim2018dialog, hou2022bring, schwarz2020improving, kojima2021large, futami2024phoneme, wu2024deferred, huang2023contextualized, arora2024semi}, Transformer or Conformer AED models \cite{tsunoo2019transformer, hori2020transformer, hori2021advanced, wei2022improving}, or Transformer-Transducer \cite{chang2021context}; and \textbf{2b) cross-utterance speech contexts fusion} that is achieved via either: \textbf{i)} input audio feature concatenation \cite{hori2020transformer, hori2021advanced};  \textbf{ii)} concatenation of cross-utterance Encoder embeddings~\cite{hori2020transformer, hori2021advanced}; or \textbf{iii)} pooling projection of cross-utterance Encoder embeddings \cite{tsunoo2019transformer, cui23_interspeech}.

\vspace{-0.4CM}
\subsection{Key Research Problems and Methodology Design}
\label{intro_problems}
Efforts to derive effective and efficient approaches to model cross-utterance speech contexts for current ASR systems based on Conformer-Transducer (C-T) 
face several challenges:

\textbf{1) Limited prior researches on cross-utterance speech contexts modeling for C-T ASR models}. Previous researches mainly targeted non-Conformer-Transducer based E2E ASR architectures, for example, RNN-based models \cite{chang2023context, gong2023longfnt, gong2024advanced, kim2019gated, kim2018dialog, hou2022bring, schwarz2020improving, kojima2021large}, Transformer or Conformer AED models \cite{tsunoo2019transformer, hori2020transformer, hori2021advanced, wei2022improving}, or Transformer-Transducer based models \cite{chang2021context}, while very limited prior researches \cite{cui23_interspeech} investigated cross-utterance speech contexts modeling  for Conformer-Transducer ASR systems.

\textbf{2) Lack of cross-utterance speech contexts modeling in speech foundation models}. Despite their huge success in scaling up pre-training data quantity, model complexity, task and domain generalization\cite{baevski2020wav2vec, chen2022wavlm, hsu2021hubert, pmlr-v162-baevski22a, babu2021xls, baevski2022data2vec, radford2023robust}, the speech contexts used in these systems are  largely confined to those at the single utterance level in both pre-training and fine-tuning stages.

\textbf{3) Lack of a complete study of cross-utterance speech contexts fusion methods}, as set out above in the last paragraph in Section \ref{intro_approach} from \textbf{i)} to \textbf{iii)}.  Previous researches \cite{hou2022bring, hori2020transformer, hori2021advanced, cui23_interspeech} only considered one or two fusion approaches among these, while lacking a complete and side-by-side comparison of their performance and efficiency.

\textbf{4) Lack of cross-utterance speech contexts modeling methods tailored for streaming and non-streaming ASR systems}. Previous researches in this direction \cite{hou2022bring, hori2020transformer, hori2021advanced, cui23_interspeech} made no differentiation among these two distinct types of E2E ASR architectures that operate with fundamentally different performance-efficiency trade-offs. To date, there is a lack of established principles for cross-utterance speech contexts fusion methods that are tailored for streaming and non-streaming ASR systems’ respective architecture designs.

\textbf{5) Efficient use of cross-utterance speech contexts in model training}. Batch-mode training of state-of-the-art (SOTA) non-contextual ASR systems \footnote{https://github.com/espnet/espnet/blob/master/espnet2/iterators/\\sequence\_iter\_factory.py} \cite{ott2019fairseq,yao2023zipformer,watanabe2018espnet}, including current weakly supervised or self-supervised foundation models\footnote{https://huggingface.co/facebook/hubert-base-ls960}\footnote{https://huggingface.co/facebook/wav2vec2-large-960h}\cite{
baevski2020wav2vec,chen2022wavlm,hsu2021hubert, pmlr-v162-baevski22a, babu2021xls, baevski2022data2vec, radford2023robust}, are normally implemented using batches of length sorted speech utterances of comparable duration. This minimizes the synchronization overhead between minibatches during parallel training. However, prior researches \cite{chang2023context, gong2023longfnt, gong2024advanced, kim2019gated, kim2018dialog, hou2022bring, schwarz2020improving, kojima2021large,tsunoo2019transformer, hori2020transformer, hori2021advanced, wei2022improving, chang2021context,cui23_interspeech} considered a straightforward modification of batch data presentation (single speech utterance per minibatch) by further enforcing the contiguousness of cross-utterance speech contexts between minibatches. This additional constraint renders the length based utterance sorting less effective, leading to an increase in the length disparity among minibatch data, and 
a decrease in model training speed.

In order to address the above issues, this paper presents: \textbf{1)} a complete and side-by-side comparison of four types of cross-utterance speech contexts modeling approaches for C-T ASR systems in terms of performance and efficiency; \textbf{2)} a complete validation of their efficacy in both C-T models pre-training using large quantities of typical speech data and their domain fine-tuning to low-resource elderly speech across multiple languages; \textbf{3)} a novel chunk-based approach that efficiently integrates longer range speech context using sliding windows spanning over contiguous utterances is also proposed for C-T systems, in addition to the three approaches previously studied for C-T and other E2E ASR models \cite{hou2022bring, hori2020transformer, hori2021advanced, cui23_interspeech}, as discussed in the last paragraph in Section \ref{intro_approach} from \textbf{i)} to \textbf{iii)}; \textbf{4)} the optimal forms of cross-utterance speech contexts modeling methods that are bespoke to the streaming and non-streaming ASR systems that operate with very different performance-efficiency trade-offs; and \textbf{5)} an efficient batch-training scheme that uses contiguous blocks of spliced speech utterances within each minibatch to minimize the synchronization overhead during  training while preserving cross-utterance speech contexts in a left-to-right manner.

\begin{figure*}{

    \centering
    \vspace{-0.3cm}
    \includegraphics[width=0.66\textwidth]{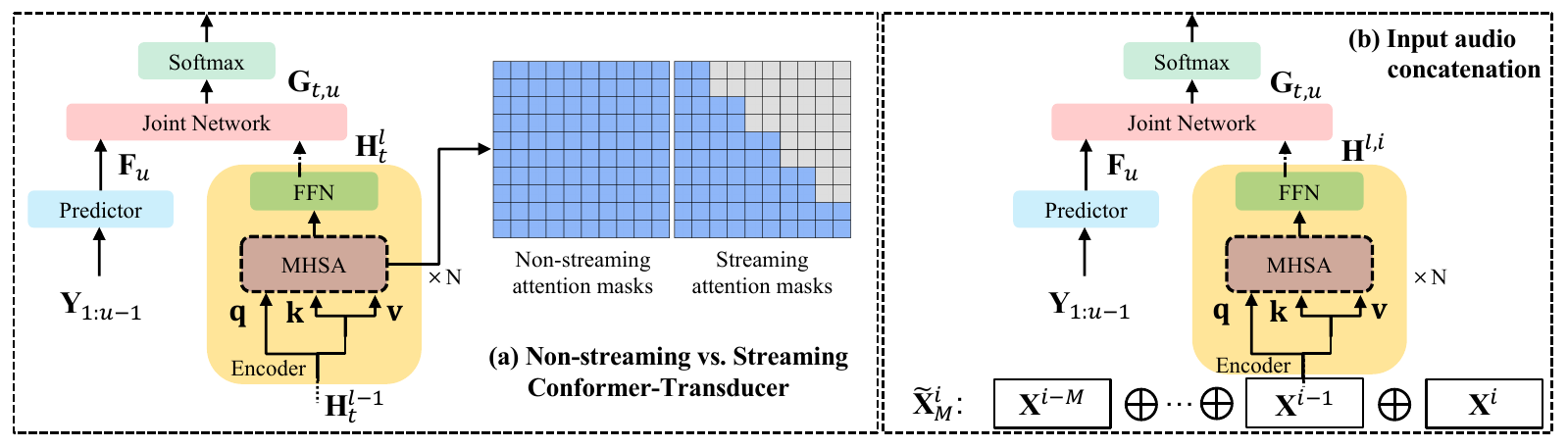}
    \vspace{-0.1cm}
    \includegraphics[width=0.66\textwidth]{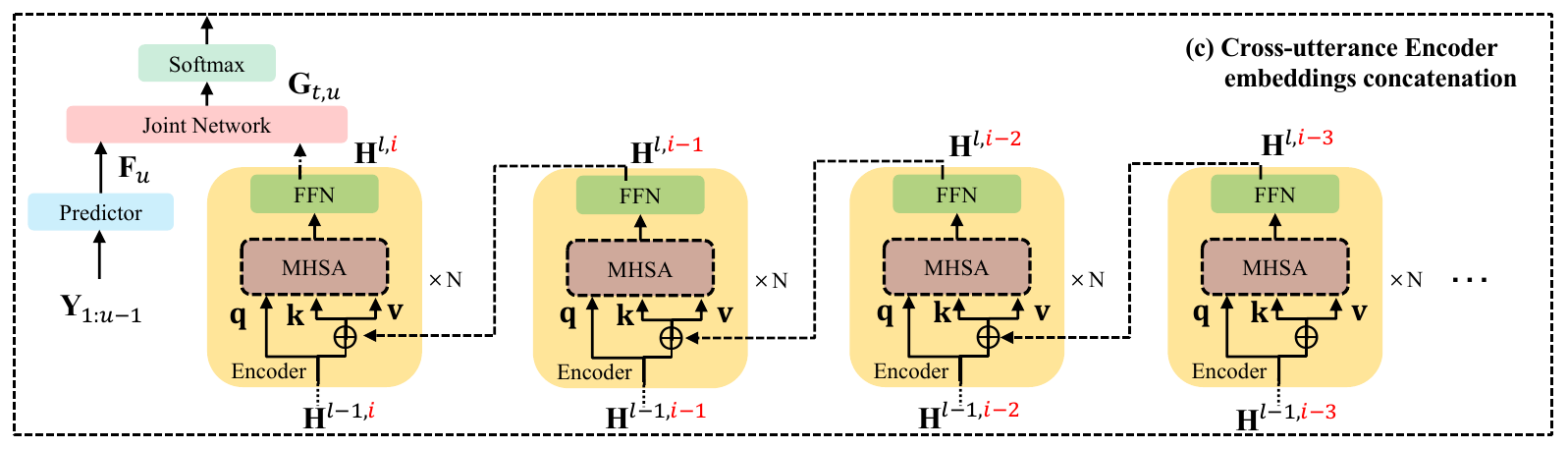}
    \vspace{-0.1cm}
    \includegraphics[width=0.66\textwidth]{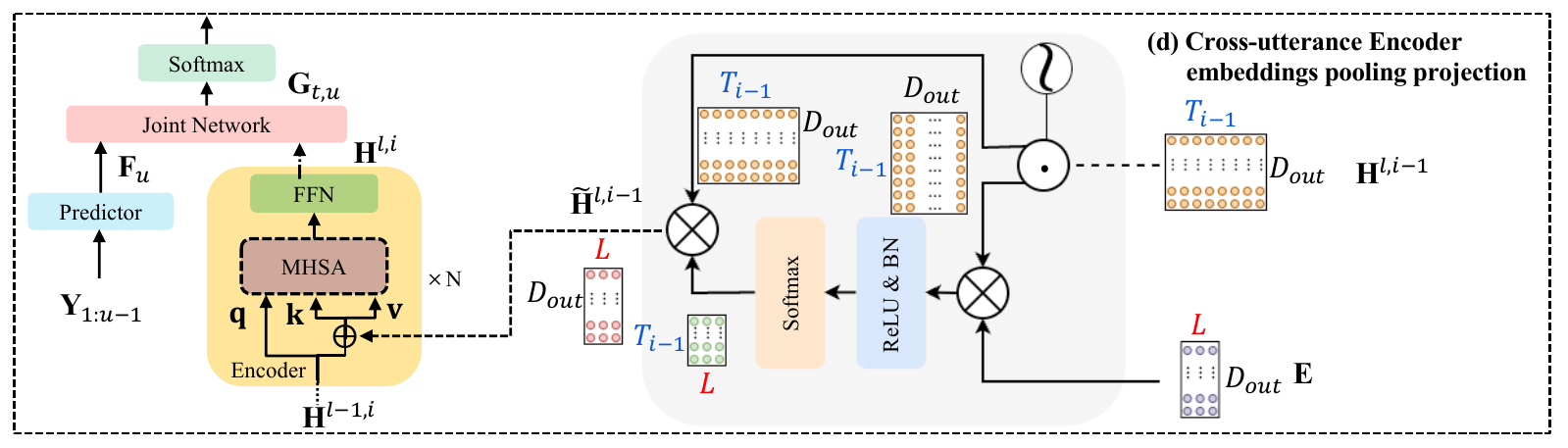}
    \vspace{-0.1cm}
    \includegraphics[width=0.66\textwidth]{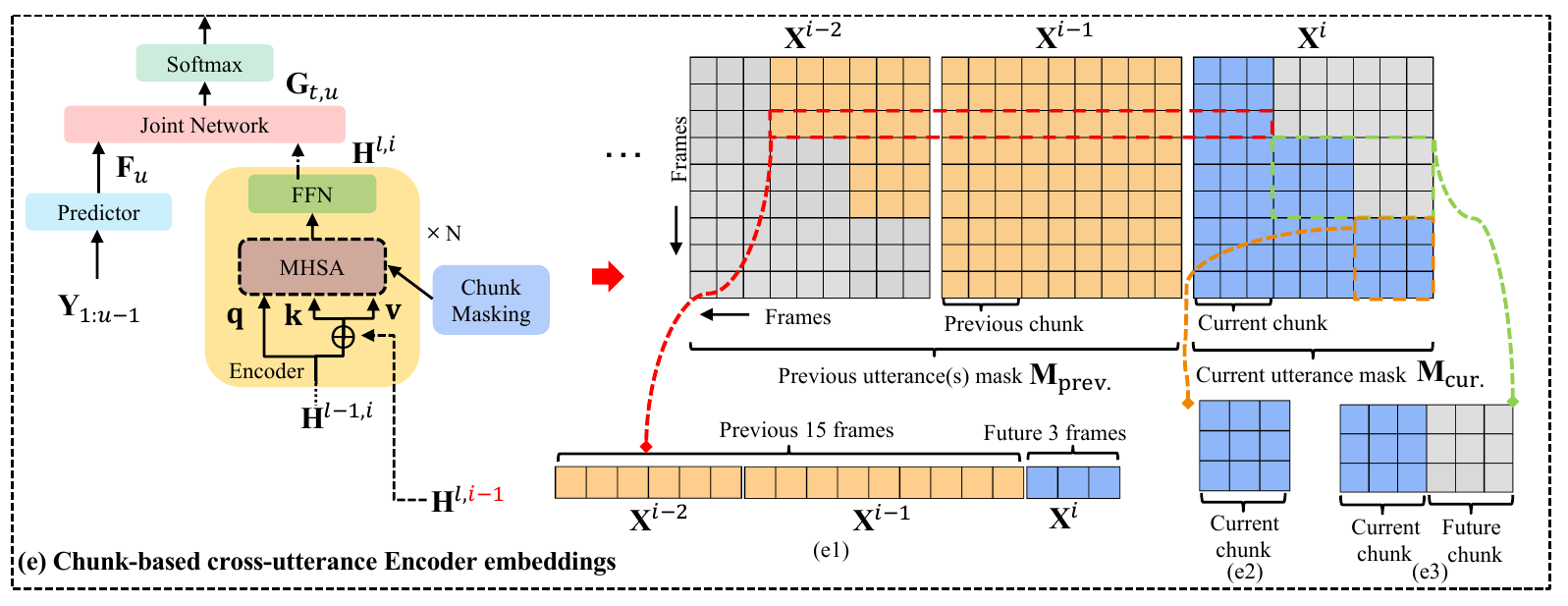}
    \caption{Example of: {\bf (a)} Standard Conformer-Transducer (C-T) models operating in 
    non-streaming or streaming mode of Section \ref{sec:streamingCT}, and also indicated by their use of full or triangular Encoder attention mask matrices to control input audio context (top corner of sub-figure \textbf{(a)}); Various C-T models modeling cross-utterance speech contexts using: {\bf (b)} input audio concatenation 
    of Section \ref{input_audio}; 
    {\bf (c)} cross-utterance Encoder embedding concatenation
    of Section \ref{cross_encoder_emb};
    {\bf (d)} cross-utterance Encoder embedding pooling projection of Section \ref{pooling_projection}; and {\bf (e)} chunk-based cross-utterance Encoder embeddings of Section \ref{chunk_based}.}
     \label{fig:rnnt}}
     \vspace{-0.5cm}
\end{figure*}

Experiments are conducted on four benchmark speech datasets across three languages: \textbf{1)} the 1000-hr English GigaSpeech \cite{chen2021gigaspeech} and Mandarin Chinese Wenetspeech \cite{zhang2022wenetspeech} speech corpora used in contextual C-T models pre-training, and \textbf{2)} the English DementiaBank Pitt elderly speech \cite{becker1994natural} and Cantonese JCCOCC MoCA \cite{xu2021speaker} elderly speech datasets used in their domain fine-tuning. 

The chunk-based cross-utterance Encoder embeddings and the cross-utterance Encoder embedding concatenation are found to be the best performing cross-utterance speech contexts modeling methods for streaming and non-streaming C-T systems, respectively. \textcolor{black}{In the pre-training domain, they consistently outperform their respective non-contextual baselines with statistically significant word error rate (WER) and character error rate (CER) reductions up to \textbf{0.9\%}, \textbf{1.1\%} absolute (\textbf{6.0\%} and \textbf{5.4\%} relative) on the GigaSpeech \textbf{Avg.} and WenetSpeech \textbf{TEST} datasets, respectively. 
These gains are accompanied by only a moderate increase in relative inference time of 3.3\% and 2.1\% on the two datasets.
} In the elderly speech domain after fine-tuning, consistent and statistically significant average WER and CER reductions up to \textbf{0.51\%} and \textbf{0.98\%} absolute (\textbf{2.0\%} and \textbf{3.4\%} relative) over their respective non-contextual baselines are also obtained on the English DementiaBank Pitt and Cantonese JCCOCC MoCA elderly speech datasets, respectively. The proposed efficient batch-mode training using utterance splicing increased the contextual C-T system training speed by up to \textbf{16.6\%} over standard training using minibatches filled with non-spliced, single utterances.

Finally, our contextual C-T systems’ performance-efficiency competitiveness against SOTA speech foundation models in both pre-training and elderly speech fine-tuning stages. In particular, on the Cantonese JCCOCC MoCA elderly speech data, our best performing contextual C-T system produced CERs comparable to those by fine-tuned Wav2vec2.0-Conformer \cite{hu2024self}, XLSR-128 \cite{hu2024self}, and Whisper-medium models (\textbf{$<$0.87\%} relative CER increase), while using less than \textbf{0.15\%} of their pre-training data quantity and \textbf{88.5\%} fewer parameters.
\vspace{-0.3CM}
\subsection{Main Contributions}
\label{contribution}
1) To the best of our knowledge, this paper presents the first work to investigate cross-utterance speech contexts modeling approaches for C-T ASR models. In contrast,  prior researches focused on non-C-T  architectures \cite{chang2023context, gong2023longfnt, gong2024advanced, kim2019gated, kim2018dialog, hou2022bring, schwarz2020improving, kojima2021large, tsunoo2019transformer, hori2020transformer, hori2021advanced, wei2022improving,chang2021context}. 

2) This paper presents the first complete study on cross-utterance speech contexts modeling for both streaming and non-streaming C-T models that are based on: \textbf{i)} input audio feature concatenation; \textbf{ii)} cross-utterance Encoder embedding concatenation; \textbf{iii)} cross-utterance Encoder embedding pooling projection; or \textbf{iv)} a novel chunk-based approach. Prior researches \cite{hori2020transformer, hori2021advanced,hou2022bring, cui23_interspeech, chen2021developing} only considered one or two approaches among the above, while lacking a complete and side-by-side comparison of their performance and efficiency.  

3) This paper empirically reveals that enforcing the consistency in previous utterance and 
current utterance contexts modeling produced the best solution for both streaming and non-streaming C-T systems. This provides insights for the practical design of C-T and other Conformer models that operate with different performance-latency trade-offs. In contrast, prior researches \cite{hori2020transformer, hori2021advanced,hou2022bring, cui23_interspeech} made no differentiation between streaming and non-streaming systems that operate with very different performance-efficiency operating points.

4) \textcolor{black}{This paper proposes an efficient batch-training scheme that uses contiguous blocks of spliced speech utterances within each minibatch. The proposed scheme minimizes the synchronization overhead during parallel training while preserving the sequential order of cross-utterance speech contexts.} In contrast, prior researches \cite{chang2023context, gong2023longfnt, gong2024advanced, kim2019gated, kim2018dialog, hou2022bring, schwarz2020improving, kojima2021large,tsunoo2019transformer, hori2020transformer, hori2021advanced, wei2022improving, chang2021context,cui23_interspeech} largely used standard batch-training with minibatches containing non-spliced, single utterances. As a result, when satisfying the contiguousness constraint of cross-utterance speech contexts between minibatches, the length disparity among minibatch data is increased, as well as the training time. Its generic nature of this batch-training scheme allows it to be applied to improve the training efficiency of both C-T and non-C-T based ASR when modeling cross-utterance speech contexts.

5) This paper conducts extensive experiments to demonstrate the efficacy of cross-utterance speech contexts empowered C-T models over non-contextual baselines in both their pre-training stage using large quantities of typical speech data and their fine-tuning to low-resource elderly speech across multiple languages. Their performance-efficiency
competitiveness against SOTA speech foundation models represented by Wav2vec2.0-Conformer \cite{hu2024self}, XLSR-128 \cite{hu2024self}, and Whisper-medium highlights the potential benefit of incorporating cross-utterance speech contexts into speech foundation models.

The rest of the paper is organized as follows: the architecture of Conformer-Transducer ASR models is reviewed in Section \ref{architecture}. Section \ref{context_representation} presents the four cross-utterance speech contexts modeling approaches for C-T systems. The proposed efficient batch-mode training scheme using utterance splicing is introduced in Section \ref{batch_mode_training}. Section \ref{experiments} presents the experimental data setup and Section \ref{experiments_results} shows the experiment results. Section \ref{conclusion} draws the conclusion and discusses future research directions. 

\vspace{-0.3cm}
\section{Conformer-Transducer ASR Architecture}
\vspace{-0.1cm}
\label{architecture}
This section provides a review of the architectures of neural Transducers, Conformer-Transducers, and their variants operating in streaming or non-streaming mode.
\vspace{-0.3cm}
\subsection{Neural Transducer}
\vspace{-0.1cm}
\label{neural_transducer}
The neural Transducer \cite{graves2012sequence} ASR model consists of three modules: an audio ``Encoder", a text ``Predictor", and a ``Joint Network", respectively, as shown in Figure \ref{fig:rnnt} (a). Let $\mathbf{X}_{1:T}=[\mathbf{x}_{1}, \mathbf{x}_{2}, ..., \mathbf{x}_{T}]$ and $\mathbf{Y}_{1:U}
=[\mathbf{y}_{1}, \mathbf{y}_{2}, ..., \mathbf{y}_{U}]$ denote the $T$ frame acoustic feature sequence and the $U$ word labels of an audio conversation, and $D_{in}$ is the Encoder's input dimensionality.
The acoustic feature sequence, $\mathbf{X}_{1:T} \in \Re^{T \times D_{in}}$, is fed into the Encoder. For each time step  $1 \leq t \leq T$, the Encoder output $\mathbf{{H}}_{t}$ is computed as follows.
\vspace{-0.2cm}
\begin{equation}
    \begin{aligned}
        \label{equ_enc}
        \mathbf{{H}}_{t} &= \mathrm{Encoder}\left(\mathbf{X}_{1:t}\right) \\ 
        \end{aligned}
\vspace{-0.2cm}
\end{equation}
The output label sequence,$\mathbf{Y}_{1:u-1}$, $1 \leq u \leq U$, is recursively fed into the Predictor to compute its representation as
\vspace{-0.2cm}
\begin{equation}
    \begin{aligned}
    \label{equ_predictor}
        \mathbf{F}_{u} &= \mathrm{Predictor}\left(\mathbf{Y}_{1:u-1}\right) \\
         \end{aligned}
         \vspace{-0.2cm}
\end{equation}
The outputs of the Encoder and Predictor are  combined in the Joint Network to obtain the hidden state $\mathbf{G}_{t, u}$ 
\vspace{-0.2cm}
\begin{equation}
    \begin{aligned}
    \label{equ_joint}
        \mathbf{G}_{t,u} &= \mathrm{JointNet}\left(\mathbf{{H}}_{t} + \mathbf{F}_{u}\right)
    \end{aligned}
    \vspace{-0.2cm}
\end{equation}
and  the Transducer output probability for the $u^{th}$ label ${\bf y}_u$,
\vspace{-0.2cm}
\begin{equation}
    \begin{aligned}
    \label{equ_relu_softmax}
        P\left(\mathbf{y}_{u}| \mathbf{X}_{1:t}, \mathbf{Y}_{1:u-1}\right) &= \mathrm{Softmax}\left(\mathbf{W}_{o} \otimes \mathbf{G}_{t, u}\right)
    \end{aligned}
    \vspace{-0.2cm}
\end{equation}
 $\mathbf{W}_{o}$ is a linear transformation applied prior to the final Softmax output layer and $\otimes$ denotes matrix multiplication. Among 
 existing neural Transducer systems, RNN, LSTM \cite{graves2012sequence, hou2022bring} and Transformer \cite{chen2021developing, zhang2020transformer, yeh2019transformer} architectures 
 are used 
 as the Encoder, while the Predictor 
 is commonly based on LSTMs. Hence, Conformer-Transducers 
 with a Conformer-based Encoder and LSTM-based Predictor are utilized in this paper. 

\vspace{-0.45cm}
\subsection{Conformer-Transducer}
\vspace{-0.1cm}
\label{conformer_transducer}
 The Conformer Encoder is based on a multi-block stacked architecture. Each block contains the following components in turn: a position wise feed-forward network (FFN) module, a multi-head self-attention (MHSA) module, a convolution (CONV) module, and a final FFN module at the end. Among these, the CONV module consists of several modules: a 1-D point-wise convolution layer, a gated linear units (GLU) activation \cite{dauphin2017language}, a second 1-D point-wise convolution layer followed by a 2-D depth-wise convolution layer, another Swish activation, and a final 1-D point-wise convolution layer. Layer normalization (LN) and residual connections are used to stabilize the training and allow more stacked layers. In the \(l^{\rm{th}}\) Conformer block ($l \geq 1$), the following operations from Equation \ref{equ_ct_ffn} to \ref{equ_ct_ln} are applied in sequence to process the previous $(l-1)^{th}$ block output $\mathbf{H}^{l-1}_t$ (By default $\mathbf{H}^0_t = \mathbf{X}_{1:t}$ denotes the input acoustic features):
 \vspace{-0.2cm}
\begin{equation}
    \begin{aligned}
    \label{equ_ct_ffn}
    {\mathbf{X}}^{l}_{t, {\rm FFN}} &= \mathbf{H}^{l-1}_t + \frac{1}{2}\mathrm{FFN}\left(\mathbf{H}^{l-1}_t\right) \\
    \end{aligned}
    \vspace{-0.2cm}
\end{equation}
The output ${\mathbf{X}}^{l}_{t, \rm FFN}$ is then used to compute the query, key, and value vectors in the MHSA module, as given by:
\vspace{-0.2cm}
\begin{equation}
\begin{aligned}
\label{equ_ct_qkv}
\mathbf{q}^{l}_t &= {\mathbf{X}}^{l}_{t, \rm  FFN}\mathbf{W}^{l}_{q} \\
\mathbf{k}^{l}_t &= {\mathbf{X}}^{l}_{t, \rm FFN}\mathbf{W}^{l}_{k} \\
\mathbf{v}^{l}_t &= {\mathbf{X}}^{l}_{t, \rm FFN}\mathbf{W}^{l}_{v}
\end{aligned}
\vspace{-0.2cm}
\end{equation}
where $\mathbf{W}^{l}_{q}$, $\mathbf{W}^{l}_{k}$ and $\mathbf{W}^{l}_{v}$ are the linear transformations to generate the query, key, and value, respectively. 
Using the above, 
the MHSA module output is computed as 
\vspace{-0.2cm}
\begin{equation}
    \begin{aligned}
    \label{equ_ct_mhsa}
            \mathbf{X}^{l}_{t, \rm MHSA} &= {\mathbf{X}}^{l}_{t, \rm FFN} + \mathrm{MHSA}\left(\mathbf{q}^{l}_t, \mathbf{k}^{l}_t, \mathbf{v}^{l}_t\right) \\
    \end{aligned}
    \vspace{-0.2cm}
\end{equation}
before being further 
processed by a CONV layer as below 
\vspace{-0.2cm}
\begin{equation}
    \begin{aligned}
    \label{equ_ct_conv}
    \mathbf{X}^{l}_{t, \rm CONV} &= \mathbf{X}^{l}_{t, \rm MHSA} + \mathrm{CONV}\left(\mathbf{X}^{l}_{t, \rm MHSA}\right) 
    \end{aligned}
    \vspace{-0.2cm}
\end{equation}
The above CONV module output is fed into further FFN and LN blocks to produce the final output $\mathbf{H}^{l}_t$, where $\mathbf{H}^{l}_t \in \Re^{T\times D_{out}}$, and $D_{out}$ is Encoder's output dimensionality
\vspace{-0.2cm}
\begin{equation}
    \begin{aligned}
       \label{equ_ct_ln}
        \mathbf{H}^{l}_t &= \mathrm{LN}\left(\mathbf{X}^{l}_{t, \rm CONV} + \frac{1}{2}\mathrm{FFN}\left(\mathbf{X}^{l}_{t, \rm CONV}\right)\right)
    \end{aligned}
    \vspace{-0.2cm}
\end{equation}


\subsection{Streaming and Non-streaming Conformer-Transducers}
\vspace{-0.1cm}
\label{sec:streamingCT}

 For practical applications requiring varying performance-latency trade-offs, non-streaming and streaming ASR systems provide different architecture designs by modeling either: {\bf 1)} the complete utterance-level context embeddings produced by, e.g. the Conformer Encoder for each speech segment being recognized;
or {\bf 2)} chunk-based partial utterance-level context representations presented in Section \ref{contribution} \textbf{iv)}. In practice, they can be
implemented using either full or triangular structured attention masks within the Conformer Encoder to control the access to the complete or partial utterance-level context. Examples of non-streaming and steaming C-T models are shown in Figure \ref{fig:rnnt} (a) (top right, left to right, respectively using full or triangular attention matrix based masks).  

\vspace{-0.3cm}
\section{cross-utterance speech contexts Representation}
\label{context_representation}
The section introduces four cross-utterance speech contexts modeling approaches for C-T systems: {\bf a)} input audio concatenation; {\bf b)} cross-utterance Encoder embedding concatenation; {\bf c)} cross-utterance Encoder embedding pooling projection; and {\bf d)} chunk-based cross-utterance Encoder embeddings.

\vspace{-0.3cm}
\subsection{Input Audio Concatenation}
\vspace{-0.1cm}
\label{input_audio}
This most straightforward approach
\cite{hori2020transformer, hori2021advanced} 
augments the current $i^{th}$ speech utterance $\mathbf{X}^{i}$ input features (with frame indices $1 \leq t \leq T_i$ omitted for brevity in this section) by its most recent $M$ preceding utterances as
\vspace{-0.2cm}
\begin{equation}
    \begin{aligned}
    {\Tilde{\mathbf{X}}}^{i}_M &= \mathbf{X}^{i-M}\oplus \cdots \oplus  \mathbf{X}^{i-1} \oplus  \mathbf{X}^{i}
        \end{aligned}
    \label{input_feature}
    \vspace{-0.2cm}
\end{equation}
before being fed into the Conformer Encoder input layer. Here, $\oplus$ denotes matrix concatenation. Except the Encoder input dimensionality needs to be enlarged, no further modification is required to the standard Conformer Encoder of Figure~\ref{fig:rnnt} (a).
An example C-T model using input audio concatenation is shown in Figure \ref{fig:rnnt} (b).


\vspace{-0.5cm}
\subsection{Cross-utterance Encoder Embedding Concatenation}
\vspace{-0.1cm}
\label{cross_encoder_emb}

Alternatively, the Conformer Encoder embeddings computed using the current, and also most recent preceding speech utterances~\cite{cui23_interspeech} can be concatenated to serve as longer span context representations. 
Given the current speech utterance $\mathbf{X}^{i}$,
 the C-T model's $l^{\rm{th}}$ Encoder layer embedding 
 is computed as Equation \ref{equ_ct_ffn} to obtain $\mathbf{X}^{l, i}_{\rm FFN}$.
These are being augmented with the Encoder embeddings computed from the previous \((i-1)^{\rm{th}}\) utterances as 
\vspace{-0.2cm}
\begin{equation}
    \begin{aligned}
    \label{equ_emb_concat}
    \mathbf{\hat{X}}^{l,i} &= \mathbf{X}^{l,i}_{\rm FFN} \oplus \mathrm{SG}\left(\mathbf{H}^{l,i-1}\right) 
    \end{aligned}
    \vspace{-0.2cm}
\end{equation} 
where $\mathbf{H}^{l,i-1} \in \Re^{(T_{i-1}) \times D_{out}}$ is the $l^{th}$ Conformer block's output for the previous \((i-1)^{{th}}\) utterance computed using Equation \ref{equ_ct_ffn} to \ref{equ_ct_ln}.
SG(·) denotes the “stop gradient” operator, and $\oplus$ is the matrix concatenation operation. 

The cross-utterance speech contexts representation $\mathbf{\hat{X}}^{l,i}$ obtained in Equation \ref{equ_emb_concat} 
is used to compute the {\textit{\textbf{contextual}}} key and value vectors in Equation~\ref{equ_ct_qkv}, while the embedding ${\mathbf{X}}^{l, i}_{\rm FFN}$ calculated from Equation \ref{equ_ct_ffn} is used to compute the query vector.
An example of C-T model using cross-utterance Encoder embedding concatenation is shown in Figure \ref{fig:rnnt} (c). 


\begin{figure*}[htbp]
    \centering
    \vspace{-1mm}
    \setlength{\abovecaptionskip}{-0cm}
    \centering
    \includegraphics[width=5.8in]{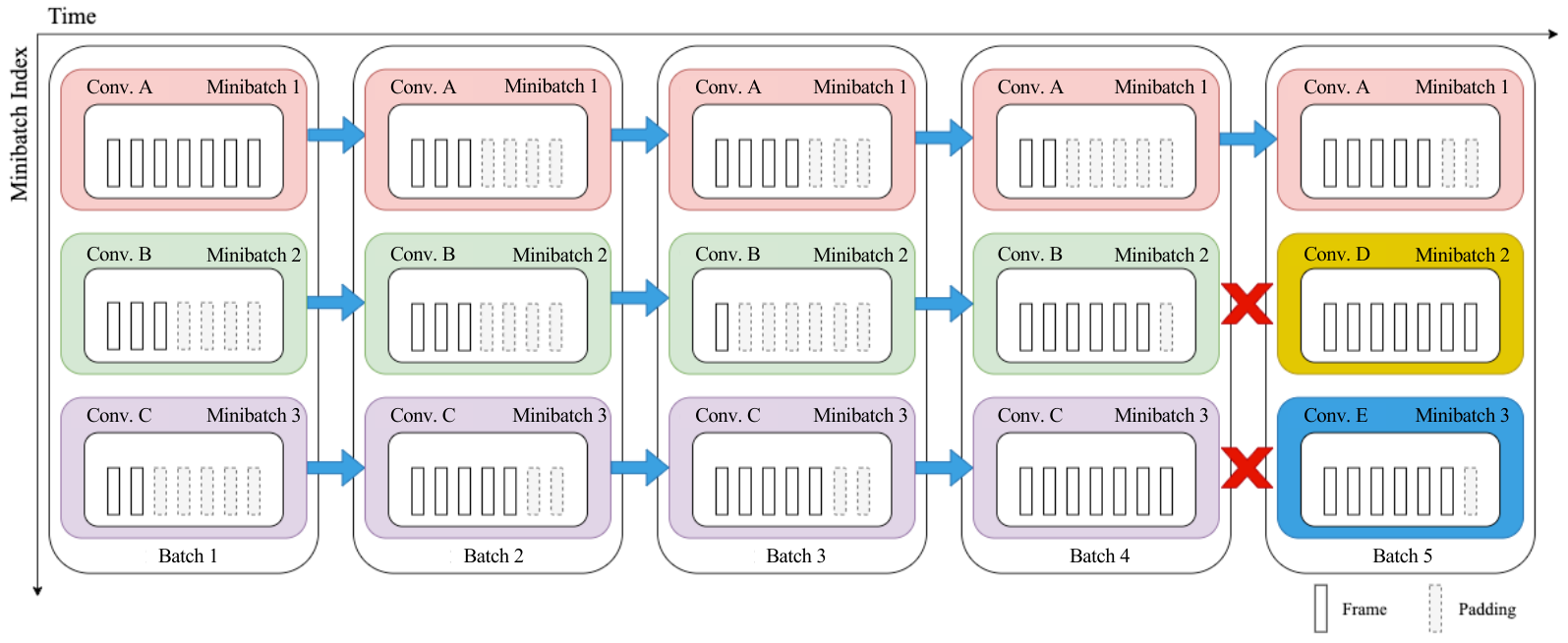}
    \vspace{-0.3cm}
    \includegraphics[width=5.8in]{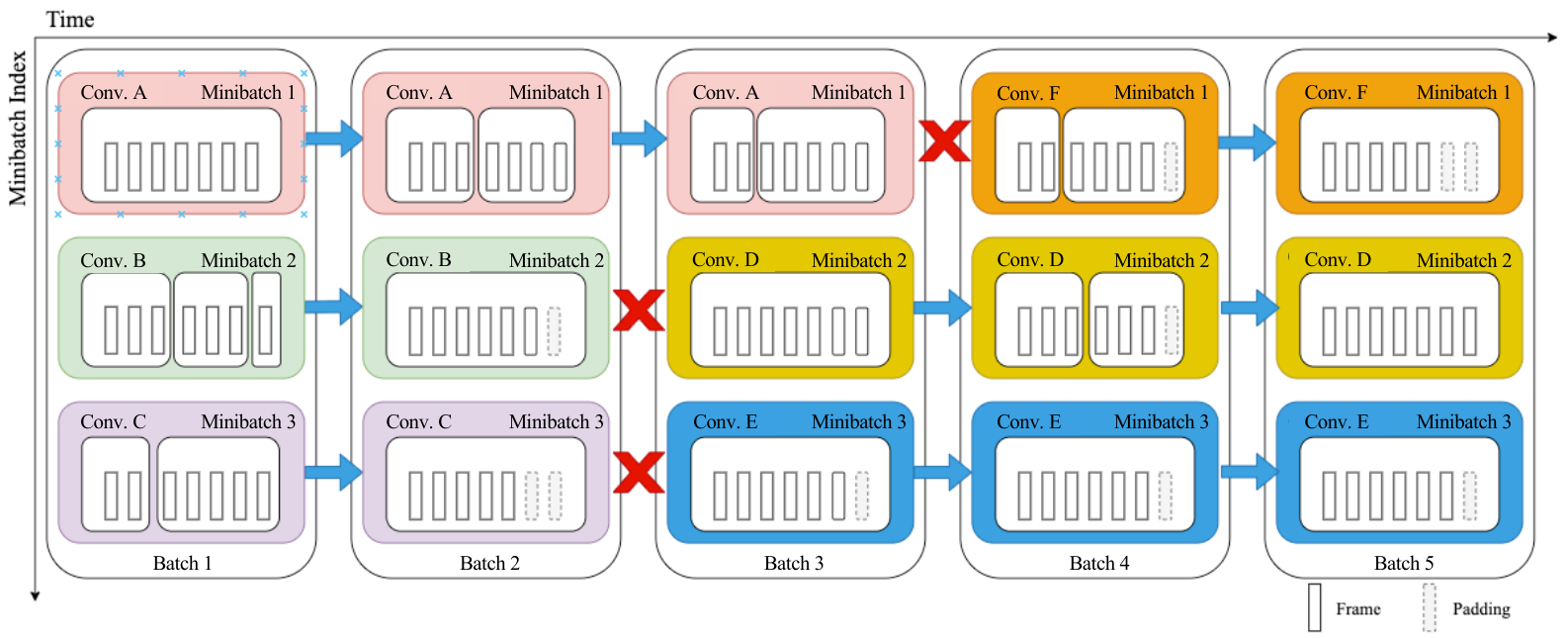}
    \caption{Examples of data serialization during batch-mode training of contextual C-T systems without (top) or with (bottom) neighbouring utterances splicing in minibatches \textcolor{black}{are shown}. In both cases, the GPU memory holds a total of 3 minibatches, each of which stores a maximum of 7 frames of audio data, e.g. shown as 7 small rectangular blocks within a larger block for minibatch 1 of Batch 1 processing one speech utterance of conversation (“Conv.") A in the top left corner of both sub-figures. The GPU memory is filled 5 times to process serialized speech utterances within a conversation according to utterances’ starting times. Using standard batch-mode training by \textbf{filling each minibatch with one single utterance}, while preserving to the left-to-right order of cross-utterance speech contexts, a GPU memory utilization rate of  63.8\% (67 out of $3 \times 7 \times 5 = 105$ frames) is achieved. In contrast, a higher GPU memory utilization rate of  90.4\% (95 out of 105 frames) is achieved by \textbf{splicing multiple neighbouring utterances within minibatches} to minimize the synchronization overhead during training. \textcolor{black}{Across the boundaries between the last and the first utterances of different speech recordings (e.g. the last utterance of conversation A in minibatch 1 of batch 3 marked in pink,  and the first utterance of conversation F in minibatch 1 of batch 4 marked in orange 
    (bottom)), cross-utterance speech contexts propagation is reset to null during contextual C-T model training (and also evaluation).} This is indicated by red crosses in the bottom sub-figure, while the blue arrows denote cross-utterance speech contexts propagation within the same speech recording (e.g. all the utterances of conversation B marked in orange used to continuously fill minibatch 2 of both batch 1 and batch 2). “Conv.” \textcolor{black}{is} short for conversation.}
 
    \label{fig:dataloader}
    \vspace{-0.3cm}
\end{figure*}

\vspace{-0.5cm}
\subsection{Cross-utterance Encoder Embedding Pooling Projection}
\vspace{-0.1cm}
\label{pooling_projection}
The cross-utterance Encoder embedding concatenation approach 
of Section \ref{cross_encoder_emb} uses high dimensional Encoder representations, as the results of the context expansion operation in Equation \ref{equ_emb_concat}. This not only incurs further computational overhead during inference but also lacks a targeted approach to locate the most useful parts of preceding utterance contexts. 

To address these issues, a more compact, lower dimensional pooling projection of Encoder embeddings is used. For the most recent  $(i-1)^{{th}}$ utterance, its Encoder embeddings at the $l^{th}$ layer,
$\mathbf{H}^{l,i-1} \in \Re^{T_{i-1}\times D_{out}}$,
are  attention-pooled~\cite{santos2016attentive} and projected to lower dimensional vectors  
$\mathbf{\Tilde{H}}^{l,i-1} \in \Re^{L \times D_{out}}$ as
\vspace{-0.2cm}
\begin{eqnarray}
    \label{eqn:pooling}
        \mathbf{\Tilde{H}}^{l,i-1} &=&
        \hspace{6.8cm}
        \\ 
        && 
        \!\!\!\! \!\!\!\! \!\!\!\! \!\!\!\! \!\!\!\! 
        \mathrm{Softmax} \left(\mathrm{BN}\left(\mathrm{ReLU}\left(\mathbf{E} 
        \otimes \mathrm{SG}\left(\mathbf{H}^{l,i-1}\right)^{\top}\right)\right)\right)\otimes {\mathbf{H}}^{l, i-1}
        \nonumber
\end{eqnarray}
where \(L \ll T_{i-1}\), $\mathbf{E} \in \Re^{L \times D_{out}}$ is a low-rank projection matrix to be learned, $\mathrm{BN}(\cdot)$ stands for Batch Normalization, and $\otimes$ denotes matrix multiplication. 

The resulting more compact $L \times D_{out}$ 
preceding utterance's embeddings then used to augment those of the current utterance in Equation \ref{equ_emb_concat}, replacing ${\mathbf{H}}^{l, i-1}$,
before being further used to compute the key and value vectors using Equation \ref{equ_ct_qkv}.
An example of C-T models using such cross-utterance Encoder embedding pooling projection
is shown in Figure \ref{fig:rnnt} (d).

\vspace{-0.3cm}
\subsection{Chunk-based Cross-utterance Encoder Embeddings}
\vspace{-0.1cm}
\label{chunk_based}
The chunk-based cross-utterance speech contexts modeling approach extends the conventional streaming ASR systems described in Section \ref{sec:streamingCT} by incorporating partial utterance representations over fixed-size context windows~\cite{chen2021developing}. This approach employs the
cross-utterance Encoder embedding concatenation operation in Equation \ref{equ_emb_concat}, while further introducing truncated, chunk-based access to the current and previous utterances’ Encoder
embeddings. The MHSA from Equation \ref{equ_ct_mhsa} is modified as follows,
\vspace{-0.2cm}
\begin{equation}
    \begin{aligned}
    \label{equ_chunk_mhsa}
            \mathbf{X}^{l}_{ \rm MHSA} &= {\mathbf{X}}^{l}_{\rm FFN} + \mathrm{MHSA}\left(\mathbf{q}^{l}, \mathbf{k}^{l}, \mathbf{v}^{l}, 
            \mathbf{M}^{l}_{\rm prev.}, \mathbf{M}^{l}_{\rm cur.}\right) \\
    \end{aligned}
    \vspace{-0.2cm}
\end{equation}
where $\mathbf{M}^{l}_{\rm prev}$ and $\mathbf{M}^{l}_{\rm cur}$ are the  mask matrices respectively controlling the
access to the current and previous utterances’ context embeddings in the $l^{\text{th}}$ Conformer block as follows. An example of C-T model using such chunk-based cross-utterance Encoder embeddings is also shown in Figure \ref{fig:rnnt} (e).

\textbf{1) Previous and current utterances’ context spans} are determined by their accumulated chunk (i.e. context widow) sizes. For example, the $1^{st}$ frame of the $i^{th}$ current speech utterance $\mathbf{X}^{i}$ (Figure \ref{fig:rnnt} (e), red arrow), which falls within its 3-frame long chunk window (Figure \ref{fig:rnnt} (e1), top right, 3 blue mask matrix elements circled in dotted red lines, all filled with the value of 1), can access up to 9 and 6 frames of context embeddings that are respectively computed from the two previous utterances, $\mathbf{X}^{i-1}$ and $\mathbf{X}^{i-2}$ (Figure \ref{fig:rnnt} (e1), top middle, 9 + 6 = 15 frames in total, indicated as yellow mask matrix elements circled in dotted red lines, filled with 1). By default, for the $1^{st}$ frame of the $i^{th}$ current speech utterance, the available current utterance’s left context span is 0. 

\textbf{2) Within-chunk context access} is fully allowed for any frames that fall within the same
chunk. For example, when processing any of the 3 frames within the $2^{nd}$ chunk for the current
utterance $\mathbf{X}^{i}$, each can access the context embeddings of the other two frames within the same chunk (Figure \ref{fig:rnnt} (e2), blue coloured mask submatrix of 3 $\times$ 3 in size, each element circled in dotted orange lines, filled with 1).

\textbf{3) Between-chunk context access} is forbidden for any two frames that are in different chunks. For example, when processing any of the 3 frames within the $2^{nd}$ chunk for the current utterance $\mathbf{X}^{i}$, the access to any of the context embeddings in the $3^{rd}$ chunk is not allowed (Figure \ref{fig:rnnt} (e3), grey coloured mask submatrix of 3 $\times$ 3 in size, each element circled in dotted orange lines, filled with 0).

\vspace{-0.3cm}
\section{Efficient batch-mode Training Using Utterance Splicing}
\label{batch_mode_training}
To enhance the training efficiency for C-T systems using cross-utterance speech contexts, an efficient batch-mode training scheme using utterance splicing is presented in this section. 
\vspace{-0.7cm}
\subsection{Cross-utterance Data Serialization in batch-mode Training}
\vspace{-0.1cm}
\label{standard_batch_mode}
To preserve the time sequence order of neighbouring speech utterances, for example, within a conversation, it is necessary to perform data serialization based on utterances’ start times during batch-mode training of C-T models that utilize cross-utterance speech contexts. In order to balance the loading of data and to minimize the synchronization overhead during training, when applying the above form of batch-mode training to SOTA non-contextual ASR systems \cite{ott2019fairseq, watanabe2018espnet, yao2023zipformer}, including the mainstream weakly supervised or self-supervised foundation models \cite{baevski2020wav2vec, chen2022wavlm, hsu2021hubert, pmlr-v162-baevski22a, babu2021xls, baevski2022data2vec, radford2023robust}, training efficiency can be further enhanced by filling minibatches with length sorted speech utterances of comparable duration.  However, straightforward application of this form of utterance length sorting based batch-mode training to contextual ASR models \cite{chang2023context, gong2023longfnt, gong2024advanced, kim2019gated, kim2018dialog, hou2022bring, schwarz2020improving, kojima2021large,tsunoo2019transformer, hori2020transformer, hori2021advanced, wei2022improving, chang2021context,cui23_interspeech} is problematic. This is because further enforcing the contiguousness of cross-utterance speech contexts between successive minibatches over time renders the length based utterance sorting less effective. This leads to an increase in the length disparity among minibatch data and at the same time a decrease in model training speed. 

An example of data serialization during batch-mode training of contextual C-T systems is shown in Figure \ref{fig:dataloader} (top). In this example, the GPU memory holds a total of 3 minibatches, each 
storing a maximum 7-frame data sequence (marked in small rectangular blocks). The GPU memory is filled 5 times to process serialized speech utterances within a conversation according to 
their starting times. Using standard batch-mode training by \textbf{filling each minibatch with one single utterance}, while preserving 
the left-to-right order of cross-utterance speech contexts, a relatively low GPU memory utilization rate of 63.8\% (67 out of $3 \times 7 \times 5 = 105$ frames) is achieved.
\vspace{-0.3cm}
\subsection{Cross-utterance Data Serialization with Utterance Splicing}
\vspace{-0.1cm}
\label{utterance_splicing}
To this end, this paper proposes an efficient batch-mode training scheme that fills contiguous blocks of spliced speech utterances into each minibatch to minimize the synchronization overhead during parallel training, while preserving cross-utterance speech contexts in a left-to-right monotonic manner. An example of data serialization during batch-mode training of contextual C-T systems with neighbouring utterances splicing in minibatches is shown in Figure \ref{fig:dataloader} (bottom). In contrast to the standard batch-mode training of Section \ref{standard_batch_mode} (also in Figure \ref{fig:dataloader} (top)) by filling each minibatch with one single utterance, a much higher GPU memory utilization rate of  90.4\% (95 out of 105 frames) is achieved by \textbf{splicing multiple neighbouring utterances within minibatches} to minimizes the synchronization overhead during batch-mode training.

The boundaries between the last and the first utterances of different speech recordings (e.g. the last utterance of conversation A in minibatch 1 of batch 3 marked in pink,  and the first utterance of conversation F in minibatch 1 of batch 4 in orange of Figure \ref{fig:dataloader} (bottom)) require cross-utterance speech contexts propagation to be reset to null during contextual C-T model training (and also evaluation). This is indicated by red crosses in the bottom sub-figure, while the blue arrows denote cross-utterance speech contexts propagation within the same speech recording (e.g. all the utterances of conversation B marked in orange used to continuously fill minibatch 2 of both batch 1 and batch 2). “Conv.” is in short for conversation.

This generic form of batch-training scheme using spliced neighbouring utterances can also be applied to improve the training efficiency of a wide range of non-C-T based ASR systems when modeling cross-utterance speech contexts.

\vspace{-0.3cm}
\section{Experiment Setup}

\label{experiments}
This section is organized as follows: Section \ref{task_description} describes four benchmark speech datasets across three languages: {\bf 1)} the widely used 1000-hr English GigaSpeech M-size \cite{chen2021gigaspeech} and 1000-hr Mandarin Chinese WenetSpeech M-size \cite{zhang2022wenetspeech} corpora used for contextual C-T pre-training, and {\bf 2)} the 
English DementiaBank Pitt and  Cantonese JCCOCC MoCA \cite{xu2021speaker} elderly speech datasets for their domain fine-tuning. The following Section \ref{experimental_setup_c_t} presents the details of the non-streaming and streaming C-T configurations, while Section \ref{experiments_elderly_setup} describes the fine-tuning setups on elderly speech.


\vspace{-0.5cm}
\subsection{Task Descriptions}
\vspace{-0.1cm}
\label{task_description}
\subsubsection{\textbf{Pre-training Datasets}}

\textbf{a)} \textit{The English GigaSpeech Corpus} is a large-scale, multi-domain English speech recognition corpus \cite{chen2021gigaspeech} with 10,000-hour high quality labeled audio collected from audiobooks, podcasts, and YouTube, encompassing both reading and spontaneous speaking styles. It spans a wide range of subject areas, including arts, science, sports, technology, education, entertainment, health, and business. In this paper, the GigaSpeech M-size corpus with 1000-hour speech is used for pre-training. Evaluation is conducted on the standard 12-hour \textbf{DEV} and 40-hour \textbf{TEST} sets.

\textbf{b)} \textit{The Mandarin WenetSpeech Corpus} is a multi-domain corpus \cite{zhang2022wenetspeech} consisting of over 10,000 hours of labeled speech from YouTube and podcasts, covering a variety of speaking styles, scenarios, domains, topics, and noisy conditions. In this paper, the WenetSpeech M-size corpus with 1000-hour speech is utilized for pre-training. A 15-hour manually labeled high-quality dataset from real meetings, i.e., Test\_Meeting, is used as the \textbf{TEST} set, featuring natural and realistic meeting room conversations sampled from 197 meetings. The topics covered include education, finance, technology, and interviews.

\subsubsection{\textbf{Elderly Speech Datasets}}
\textbf{a)} \textit{The English DementiaBank Pitt Corpus} contains 33 hours of audio from 292 AD assessment interviews between elderly participants (\textbf{PAR.}) and clinical investigators (\textbf{INV.}) \cite{becker1994natural}. It is split into a 27.2-hour training set with 688 speakers (244 elderly, 444 investigators), a 4.8-hour development set with 119 speakers (43 elderly, 76 investigators), and a 1.1-hour evaluation set with 95 speakers (48 elderly, 47 investigators)\footnote{The evaluation set includes Cookie Theft task recordings from 48 speakers following the ADReSS challenge \cite{luz2020alzheimer}, while the development set contains other task recordings from these same speakers if available.}.  There is no speaker overlap between the training and the development or evaluation sets. Silence stripping and data augmentation (using both speaker-independent and elderly speaker-dependent speed perturbation \cite{ye2021development}) lead to a 58.9-hour augmented training set, a 2.5-hour \textbf{DEV} set, and a 0.6-hour \textbf{EVAL} set. 

\textbf{b)} \textit{The Cantonese JCCOCC MoCA Corpus} consists of 256 cognitive assessment interviews between elderly participants and clinical investigators \cite{xuspeaker}. It is divided into a training set with 369 speakers (158 elderly, 211 investigators), and development and evaluation sets each with 49 elderly speakers not in the training set. Silence stripping and data augmentation~\cite{geng2022speaker} further lead to a 156.9-hour augmented training set, a 3.5-hour \textbf{DEV} set, and a 3.4-hour \textbf{EVAL} set.

\begin{table*}[t]
\caption{Performance (WER/CER\%) of different cross-utterance speech contexts fusion approaches for non-streaming C-T models. The evaluation is conducted on the standard \textbf{dev} and \textbf{test} sets of the GigaSpeech M-size dataset, and the \textbf{test} set of the WenetSpeech M-size pre-training corpus. Here (A) stands for input audio concatenation in Section \ref{input_audio}, (B) is short for cross-utterance encoder embedding concatenation of Section \ref{cross_encoder_emb}, (C) stands for cross-utterance encoder embedding pooling projection of Section  \ref{pooling_projection}\protect\footnotemark, and (D) is chunk-based cross-utterance Encoder embeddings of Section \ref{chunk_based}. ``$\dagger$'' and ``$*$'' denote a statistically significant (MAPSSWE\cite{gillick1989some}, $\alpha=0.05$) difference over the baseline systems (Sys.1, 13). ``utt.'', ``Uttlen.'', ``Avg.'', and ``ctx.'' are respectively short for utterance, utterance length, average and context
}.
\vspace{-0.5cm}
\centering
\setlength{\tabcolsep}{8.5pt} 
\fontsize{8}{9.5}\selectfont 
\begin{tabular}{c|c|c|c|ccc|c|c|c|c|c} 
\hline\hline
\multirow{3}{*}{ID} & 
Batch&
\multirow{2}{*}{Fusion} &
\multirow{2}{*}{\#Previous utt.} &
\multicolumn{5}{c|}{GigaSpeech} &
\multicolumn{3}{c}{WenetSpeech} \\
\cline{5-12}
&mode & \multirow{2}{*}{methods} & \multirow{2}{*}{ctx. length} & \multicolumn{3}{c|}{WER} & Training & \multirow{2}{*}{RTF} & CER & Training & \multirow{2}{*}{RTF} \\
\cline{5-7}
\cline{10-10}
&training&&&DEV&TEST&Avg.& time &&TEST&time \\
\hline\hline
1 &  & - & - & 15.0 & 14.9 & 14.9& 40.1h & 0.0243 & 20.48 & 38.1h & 0.0241\\
\cline{3-12}
2 & & (A) & 1 $\times$ Uttlen. & 15.2 & 15.1 &15.1& 48.5h & 0.0257 & 20.72 & 46.3h & 0.0255\\
\cline{3-12}
3 &  &  & 1 $\times$ Uttlen. & $14.2^\dagger$ & $14.1^\dagger$ & $14.1^\dagger$ &47.2h & 0.0254 & $19.85^\dagger$ & 45.8h & 0.0250\\
4 &   & (B) & 2 $\times$ Uttlen. & $\textbf{14.1}^\dagger$ & $14.1^\dagger$ &$14.1^\dagger$ &  54.5h & 0.0257 & $19.45^\dagger$ & 52.8h & 0.0254\\
5 &   &  & 3 $\times$ Uttlen. & $\textbf{14.1}^\dagger$ & $\textbf{14.0}^\dagger$ & $\textbf{14.0}^\dagger$ &59.6h & 0.0258 & $\textbf{19.38}^\dagger$ & 57.4h & 0.0256\\
\cline{3-12}
6 &  Without&  & 1 $\times$ 32 & 14.8 & 14.7 &14.7& 41.5h & 0.0251 & 20.28 & 40.0h & 0.0246\\
7 & utterance& (C) & 2 $\times$ 32 & 14.6 & 14.6&14.6 & 43.7h & 0.0253 & 20.20 & 42.2h & 0.0250\\
8 &  splicing&  & 3 $\times$ 32 &$14.6^\dagger$ & $14.5^\dagger$ & $14.5^\dagger$&46.2h & 0.0254 & $20.14^\dagger$ & 45.3h & 0.0245\\
\cline{3-12}
9 &  & & $\leq$60 frames & $14.6^\dagger$ & $14.6^\dagger$ & $14.6^\dagger$ &42.3h & 0.0252 & 20.23 & 41.4h & 0.0245\\
10 &   &  \multirow{2}{*}{(D)}& $\leq$100 frames & $14.5^\dagger$ & $14.5^\dagger$ & $14.5^\dagger$ &44.9h & 0.0253 & $20.04^\dagger$ & 44.1h & 0.0252\\
11 & &   & $\leq$160 frames& $14.4^\dagger$ & $14.4^\dagger$ & $14.4^\dagger$ &49.7h & 0.0256 & $19.90^\dagger$ & 48.6h & 0.0252\\
12 & & & $\leq$200 frames & $14.4^\dagger$ & $14.4^\dagger$ & $14.4^\dagger$ &53.1h & 0.0257 & $19.80^\dagger$ & 52.0h & 0.0254\\
\cline{2-12}
13  & & - & - & 15.0 & 14.9 & 14.9 &33.5h & 0.0235 & 20.45 & 32.2h & 0.0233\\
\cline{3-12}
14 &  & (A) & 1 $\times$ Uttlen. & 15.3 & 15.2 &  15.2 &41.6h & 0.0253 & 20.78 & 39.2h & 0.0251\\
\cline{3-12}
15 & &  & 1 $\times$ Uttlen. & $14.2^*$ & $14.2^*$ & $14.2^*$ &41.4h & 0.0245 & $19.82^*$ & 41.2h & 0.0244\\
16 & & (B) & 2 $\times$ Uttlen. & $\textbf{14.1}^*$ & $14.1^*$ & $14.1^*$ &48.3h & 0.0253 & $19.47^*$ & 48.1h & 0.0252\\
17 & With & & 3 $\times$ Uttlen. & $\textbf{14.1}^*$ & $\textbf{14.0}^*$ & $\textbf{14.0}^*$ & 53.4h & 0.0256 & $\textbf{19.37}^*$ & 53.1h & 0.0254\\
\cline{3-12}
18 & utterance &  & 1 $\times$ 32 & 14.7 & 14.7 & 14.7 & 34.6h & 0.0236 & 20.25 & 34.3h & 0.0235\\
19 & splicing& (C) & 2 $\times$ 32 & 14.6 & 14.6 & 14.6 & 36.8h & 0.0238 & $20.19^*$ & 36.5h & 0.0238\\
20 & &  & 3 $\times$ 32 & $14.5^*$ & $14.5^*$ &$14.5^*$ &  40.5h & 0.0244 & $20.12^*$ & 40.3h & 0.0243\\
\cline{3-12}

21 &  & & $\leq$60 frames & $14.5^*$ & $14.5^*$ & $14.5^*$ &36.5h & 0.0238 & 20.16 & 36.6h & 0.0238\\
22 &  & \multirow{2}{*}{(D)} & $\leq$100 frames & $14.5^*$ & $14.4^*$ & $14.4^*$ &38.2h & 0.0240 & $20.02^*$ & 38.0h & 0.0241\\
23 &  &  & $\leq$160 frames & $14.4^*$ & $14.4^*$ &  $14.4^*$ & 42.6h & 0.0246 & $19.93^*$ & 42.3h & 0.0246\\
24 &  & & $\leq$200 frames & $14.4^*$ & $14.4^*$ & $14.4^*$ &47.9h & 0.0252 & $19.78^*$ & 47.6h & 0.0251\\
\hline\hline
\end{tabular}
\vspace{-0.1cm}
\label{exp: giga_nonstreaming}
\end{table*}

\begin{table*}[t]
\caption{Performance (WER/CER\%) of different cross-utterance speech contexts modeling approaches for Streaming C-T models. ``$\dagger$'' and ``$*$'' denote a statistically significant (MAPSSWE\cite{gillick1989some}, $\alpha=0.05$) difference over the baseline systems (Sys.1, 17). Other notations follow Table~\ref{exp: giga_nonstreaming}.
}
\vspace{-0.3cm}
\centering
\setlength{\tabcolsep}{4.5pt} 
\fontsize{8}{9.5}\selectfont 
\begin{tabular}{c|c|c|c|c|c|ccc|c|c|c|c|c} 
\hline\hline
\multirow{3}{*}{ID} & 
Batch&
\multirow{2}{*}{Fusion} &
\multirow{2}{*}{\#Previous utt.} &
\multirow{2}{*}{\#Current utt.} &
\multirow{1}{*}{\#Future ctx.} &
\multicolumn{5}{c|}{GigaSpeech} &
\multicolumn{3}{c}{WenetSpeech} \\
\cline{7-14}
&mode&\multirow{2}{*}{method}&\multirow{2}{*}{ctx. length}&\multirow{2}{*}{ctx. length}&\multirow{1}{*}{length in }&\multicolumn{3}{c|}{WER}&Training&\multirow{2}{*}{RTF}&CER&Training&\multirow{2}{*}{RTF}\\
\cline{7-9}
\cline{12-12}
 &training &  & & & cur. utt. &DEV & TEST& Avg.&time &  & TEST & time & \\
\hline\hline
1 &  & - & - & Cur. Uttlen & \multirow{16}{*}{$\leq$20 frames} & 17.7 & 17.4 &17.5& 44.9h & 0.0291 & 24.05 & 44.0h & 0.0286\\
\cline{3-5}
\cline{7-14}
2   & & (A) & 1 $\times$ Uttlen. & Cur. Uttlen &   & 17.9 & 17.6 &17.7& 52.3h & 0.0339 & 24.32 & 50.1h & 0.0325\\
\cline{3-5}
\cline{7-14}

3  &  &  & 1 $\times$ Uttlen. & &  & $17.2^\dagger$ & $17.2^\dagger$ & $17.2^\dagger$ &51.9h & 0.0337 &$23.46^\dagger$ & 49.5h & 0.0321\\
4  &  & (B) & 2 $\times$ Uttlen. & Cur. Uttlen &   & $17.0^\dagger$ & $16.9^\dagger$ & $16.9^\dagger$ & 57.3h & 0.0372 & $23.32^\dagger$ & 55.2h & 0.0358\\
5  &&  & 3 $\times$ Uttlen. & &  & $17.0^\dagger$ & $16.9^\dagger$ &$16.9^\dagger$ & 62.5h & 0.0406 & $23.30^\dagger$ & 60.4h & 0.0392\\
\cline{3-5}
\cline{7-14}
6 & Without&  & 1 $\times$ 32 & &  & 17.6 & 17.5 & 17.5 &45.3h & 0.0298 & 23.80 & 44.4h & 0.0295\\
7 &utterance& (C) & 2 $\times$ 32 & Cur. Uttlen &   & 17.4 & 17.4 & 17.4 &47.2h & 0.0306 & 23.65 & 46.3h & 0.0301\\
8  &splicing &  & 3 $\times$ 32 & &  & $17.3^\dagger$ & $17.3^\dagger$ &  $17.3^\dagger$ &49.1h & 0.0319 & $23.50^\dagger$ & 48.2h & 0.0313\\
\cline{3-5}
\cline{7-14}
9 & & \multirow{4}{*}{-} & 0 & $\leq$60 frames&  & 17.8 & 17.5 &17.6& 44.6h & 0.0290 & 23.90 & 43.8h & 0.0284\\
10  & & & 0 & $\leq$100 frames&   & 17.7 & 17.4 &17.5& 44.8h & 0.0291 & 23.70 & 44.0h & 0.0285\\
11 & & & 0 & $\leq$160 frames&   & 17.7 & 17.4 &17.5& 45.0h & 0.0292 & 23.65 & 44.3h & 0.0288\\
12  & & & 0 & $\leq$200 frames&   & 17.7 & 17.4 &17.5& 45.0h & 0.0292 & 23.65 & 44.3h & 0.0287\\
\cline{3-5}
\cline{7-14}
13  & & & \multicolumn{2}{c|}{$\leq$60 frames}      &   & $17.1^\dagger$ & $17.0^\dagger$ & $17.0^\dagger$ & 46.7h & 0.0303 & $23.35^\dagger$ & 45.1h & 0.0296\\
14  & & \multirow{2}{*}{(D)} & \multicolumn{2}{c|}{$\leq$100 frames} &   & $17.0^\dagger$ & $16.7^\dagger$ & $16.8^\dagger$ &48.1h & 0.0312 & $23.24^\dagger$ & 46.4h & 0.0301\\
15 & &  & \multicolumn{2}{c|}{$\leq$160 frames} &   & $\textbf{16.7}^\dagger$ & $\textbf{16.4}^\dagger$ &$\textbf{16.5}^\dagger$ & 51.3h & 0.0333 & $23.10^\dagger$ & 49.5h & 0.0322\\
16  & & & \multicolumn{2}{c|}{$\leq$200 frames} &   & $\textbf{16.7}^\dagger$ & $\textbf{16.4}^\dagger$ & $\textbf{16.5}^\dagger$ &54.6h & 0.0354 & $\textbf{23.05}^\dagger$& 52.8h & 0.0343\\
\cline{2-14}
17  && - & - & Cur.Uttlen & \multirow{16}{*}{$\leq$20 frames}   & 17.7 & 17.4 & 17.5 &36.8h & 0.0254 & 24.02 & 36.1h & 0.0244\\
\cline{3-5}
\cline{7-14}
18  && (A) & 1 $\times$ Uttlen. & Cur.Uttlen &   & 17.9 & 17.7 & 17.8&44.6h & 0.0290 & 24.40 & 43.8h & 0.0284\\
\cline{3-5}
\cline{7-14}

19 & &  & 1 $\times$ Uttlen. & &   & $17.2^*$ & $17.2^*$ & $17.2^*$ &44.2h & 0.0287 & $23.44^*$ & 44.1h & 0.0286\\
20  & & (B) & 2 $\times$ Uttlen. & Cur. Uttlen &   & $17.0^*$ & $17.0^*$ & $17.0^*$ & 50.1h & 0.0325 & $23.32^*$ & 50.3h & 0.0326\\
21 &  &  & 3 $\times$ Uttlen. & &   & $17.0^*$ & $16.9^*$ & $16.9^*$ &54.3h & 0.0352 & $23.30^*$ & 54.4h & 0.0353\\
\cline{3-5}
\cline{7-14}
22  &With&  & 1 $\times$ 32 & &   & 17.5 & 17.5 & 17.5 &37.2h & 0.0262 & 23.78 & 37.1h & 0.0260\\
23 &utterance& (C) & 2 $\times$ 32 & Cur. Uttlen &   & 17.3 & 17.4 & 17.4&39.8h & 0.0269 & 23.66 & 39.5h & 0.0264\\
24  & splicing&& 3 $\times$ 32 & &   & $17.2^*$ & 17.3 & 17.3 &42.5h & 0.0281 & 23.49 & 42.4h & 0.0276\\
\cline{3-5}
\cline{7-14}
25 & & \multirow{4}{*}{-} & 0 & $\leq$60 frames&   & 17.7 & 17.5 &  17.6 &37.2h & 0.0261 & 23.92 & 36.8h & 0.0258\\
26  & & & 0 & $\leq$100 frames&   & 17.7 & 17.4&17.5 & 37.6h & 0.0264 & 23.73 & 37.0h & 0.0260\\
27 & & & 0 & $\leq$160 frames &   & 17.7 & 17.4 &17.5& 37.9h & 0.0265 & 23.67 & 37.1h & 0.0261\\
28  & & & 0 & $\leq$200 frames &   & 17.7 & 17.4&17.5 & 37.9h & 0.0265 & 23.67 & 37.1h & 0.0262\\
\cline{3-5}
\cline{7-14}
29  & & & \multicolumn{2}{c|}{$\leq$60 frames}  &   & $17.2^*$ & $17.1^*$ & $17.1^*$ &38.0h & 0.0267 & $23.40^*$ & 37.2h & 0.0261\\
30  & & \multirow{2}{*}{(D)} & \multicolumn{2}{c|}{$\leq$100 frames} &  &$17.0^*$ & $16.8^*$ & $16.9^*$ & 40.2h & 0.0275 & $23.25^*$ & 39.6h & 0.0265\\
31  & &  & \multicolumn{2}{c|}{$\leq$160 frames}  &   & $\textbf{16.8}^*$ &$\textbf{16.4}^*$& $\textbf{16.5}^*$&43.6h & 0.0283 & $23.10^*$ & 42.8h & 0.0281\\
32  & & & \multicolumn{2}{c|}{$\leq$200 frames}  &  & $\textbf{16.8}^*$ & $\textbf{16.4}^*$& $\textbf{16.5}^*$&47.8h & 0.0310 & $\textbf{23.05}^*$ & 46.5h & 0.0304\\
\cline{4-14}
\hline\hline
\end{tabular}
\vspace{-0.2cm}
\label{exp: giga_streaming}
\end{table*}

\begin{table*}[t]
\caption{Performance (WER/CER\%) contrasts between our best performing cross-utterance speech contexts conditioned C-T systems (non-streaming and streaming) and SOTA results on the English GigaSpeech M-size (GS-M) and Mandarin WenetSpeech M-size (WS-M) corpora. Here (B) denotes the cross-utterance Encoder embedding concatenation of Section \ref{cross_encoder_emb}, while (D) is the chunk-based cross-utterance Encoder embeddings of Section \ref{chunk_based}. “Enc." corresponds to Encoder, “GS-XL" indicates the GigaSpeech-XL size dataset, and “Params." represents parameters. ``C-T'' denotes ``Conformer-Transducer''.
} 
\vspace{-0.3cm}

\centering
\setlength{\tabcolsep}{2pt} 
\fontsize{8.2}{10}
\resizebox{\linewidth}{!}{
\begin{tabular}{c|c|c|c|c|c|cc|c} 
\hline\hline
\multirow{3}{*}{ID} & 
\multirow{3}{*}{System} &
\multicolumn{2}{c|}{Training data} &
\multirow{3}{*}{\#Params.} &
\multirow{2}{*}{Fusion} &
\multicolumn{2}{c|}{GigaSpeech WER} &
WenetSpeech CER \\
\cline{3-4}
\cline{7-9}
&&\multirow{2}{*}{Source} &\multirow{2}{*}{Size (\#hour)}&&\multirow{2}{*}{method}&\multirow{2}{*}{DEV} & \multirow{2}{*}{TEST} & \multirow{2}{*}{TEST} \\

&&&& & & &\\
\hline\hline
1& 2023 OpenAI Whisper-Medium  &Mixed data&680k &769M &-& 11.20 & 11.20  & 45.66\\
2& 2023 OpenAI Whisper-Medium + Lora Fine-tuning  &Mixed data + GS-M / WS-M&680k + 1k / 1k&769M &-& 9.58& 9.53  & 33.92 \\
3&2023 SJTU LongFNT\cite{gong2023longfnt}&GS-XL + GS-M& 10k + 1k &-&External Enc.& 14.80 & 14.30 & -\\ 
4&2023 SJTU LongFNT-Streaming\cite{gong2023longfnt}&GS-XL + GS-M& 10k + 1k&-&External Enc.& 19.20 & 18.20 & - \\ 
5&2021 SpeechColab Kaldi TDNN \cite{chen2021gigaspeech, zhang2022wenetspeech} &GS-M&1k&-&-& 17.96 & 17.53 & 28.22\\
6& 2024 SJTU Factorized Neural Transducer \cite{gong2024advanced}& GS-M&1k & -&-& 16.80& 16.30 & - \\
7&2023 CUHK ESPnet CFM-Transducer\cite{cui23_interspeech} &GS-M&1k&88.5M&(B) &14.30 & 14.20 & - \\ \hline
8&\textbf{Our non-streaming C-T of Section \ref{cross_encoder_emb} (Sys.17 in Table \ref{exp: giga_nonstreaming})}&GS-M / WS-M&1k / 1k&88.5M&(B)& \textbf{14.10} & \textbf{14.00} & \textbf{19.37} \\ 
9&\textbf{Our streaming C-T of Section \ref{chunk_based} (Sys.16 in Table \ref{exp: giga_streaming})}&GS-M / WS-M&1k / 1k&88.5M&(D) & \textbf{16.70} & \textbf{16.40} & \textbf{23.05} \\ 
\hline\hline
\end{tabular}}
\label{exp: existing_pre-training}
\vspace{-0.1cm}
\end{table*}

\begin{table*}[htbp]
\caption{Performance (WER/CER\%) of cross-utterance speech contexts modeling approaches for non-streaming and streaming C-T models that are fine-tuned on the elderly English DEMENTIABANK PITT and Cantonese JCCOCC MoCA elderly speech datasets. ``$\dagger$'', ``$*$'', ``$\triangle$'' and ``$\star$'' denote a statistically significant (MAPSSWE\cite{gillick1989some}, $\alpha=0.05$) difference over the baseline systems (Sys.1, 6, 11, and 16). Other notations follow Table~\ref{exp: giga_nonstreaming}.}
\vspace{-0.3cm}
\centering
\resizebox{\linewidth}{!}{
\setlength{\tabcolsep}{1.5pt} 
\fontsize{8.2}{10}\selectfont 
\begin{tabular}{c|c|c|c|c|c|c|c|cc|cc|c|c|c|c|c|c|c|c} 
\hline\hline
\multirow{4}{*}{ID} & 
\multirow{4}{*}{System} &
\multirow{2}{*}{Batch}&
 &\multirow{3}{*}{\#Previous utt.}
 &\multirow{3}{*}{\#Current utt.}
 &\multirow{1}{*}{\#Future }
 & &
\multicolumn{7}{c|}{Dementiabank Pitt WER} &
\multicolumn{5}{c}{JCCOCC MoCA CER } \\
&&\multirow{2}{*}{mode}&Fusion&\multirow{3}{*}{ctx. length} &\multirow{3}{*}{ctx. length} &\multirow{1}{*}{ ctx.}&Fine-tuned&\multicolumn{7}{c|}{(GigaSpeech $\rightarrow$ Dementiabank Pitt)}&\multicolumn{5}{c}{(WenetSpeech $\rightarrow$ JCCOCC MoCA)} \\

\cline{9-20}
& &\multirow{2}{*}{training} &method  & &&  length in&from & \multicolumn{2}{c|}{DEV} &\multicolumn{2}{c|}{EVAL} & \multirow{2}{*}{Avg.}&Fine-tuning &\multirow{2}{*}{RTF}& \multirow{2}{*}{DEV} & \multirow{2}{*}{EVAL} & \multirow{2}{*}{Avg.} & Fine-tuning &\multirow{2}{*}{RTF}\\
\cline{9-12}
& & &  & & & \multirow{1}{*}{cur. utt.}  &&    INV.& PAR. & INV. &PAR. & &time & & && &time \\
\hline\hline
1 &  &          & -   & -                  &  &  & Table \ref{exp: giga_nonstreaming}.1 & 15.41 & 35.68 & 17.59 & 25.90 &25.26 & 3.0h &0.0250& 30.10 & 27.92 & 29.01 & 8.1h & 0.0252\\
2 &  & Without  & (A) & 1 $\times$ Uttlen. & Cur. &  & Table \ref{exp: giga_nonstreaming}.2&15.80 & 36.16 & 17.94 & 26.38 & 25.70& 3.3h &0.0275& 30.28 & 28.04 & 29.16&8.9h&0.0280\\
3 &  & utterance & (B) & 3 $\times$ Uttlen. & Uttlen. & - & Table \ref{exp: giga_nonstreaming}.5 & $\textbf{15.01}$ & $\textbf{34.97}^\dagger$ & $\textbf{15.67}$ & $\textbf{25.53}$ &  $\textbf{24.70}^{\dagger}$&4.0h &0.0333& $\textbf{29.13}^{\dagger}$ & $\textbf{26.95}^{\dagger}$ & $\textbf{28.04}^{\dagger}$  & 10.1h &0.0340\\
4 &  & splicing& (C) & 3 $\times$ 32 &  &  & Table \ref{exp: giga_nonstreaming}.8 & 15.30 & 35.39 & 16.08 & 25.78 & 25.03&3.3h&0.0278& 29.92 & 27.54 & 28.73&9.0h&0.0284\\
5 & Non- & & (D) & $\leq$200 frames &  &  & Table \ref{exp: giga_nonstreaming}.12 & 15.23 & 35.10 & 15.84 & 25.65 & 24.86 &3.4h&0.0283 & 29.56 & 27.13 & 28.35 & 9.2h&0.0286\\
\cline{3-20}
6 & Streaming &   &  & - &  &  & Table \ref{exp: giga_nonstreaming}.13 & 15.32 & 35.66 & 17.38 & 25.80 & 25.19 &2.6h&0.0237 & 30.12 & 27.91 & 29.02 & 7.5h&0.0241\\
7 &  & With & (A) & 1 $\times$ Uttlen. & Cur. &  & Table \ref{exp: giga_nonstreaming}.14 & 15.70 & 36.15 & 17.73 & 26.34 & 25.64&2.8h&0.0245&30.29 & 28.05 & 29.17 & 8.2h&0.0250\\
8 &  & utterance & (B) & 3 $\times$ Uttlen. & Uttlen. & - & Table \ref{exp: giga_nonstreaming}.17 & $\textbf{14.99}$ & $\textbf{34.95}^*$ & $\textbf{15.45}$ & $\textbf{25.49}$ & $\textbf{24.68}^*$& 3.4h&0.0285& $\textbf{29.12}^{*}$ & $\textbf{26.96}^*$ & $\textbf{28.04}^*$ & 9.1h&0.0291\\
9 &  & splicing & (C) & 3 $\times$ 32 &  &  & Table \ref{exp: giga_nonstreaming}.20 & 15.29 & 35.38 & 15.97& 25.76 & 25.10 &2.8h &0.0247& 29.90 & 27.55 & 28.73 & 8.2h&0.0249\\

10 &  &  & (D) & $\leq$200 frames &  &  & Table \ref{exp: giga_nonstreaming}.24 &15.22 & 35.09 & 15.73 & 25.63 & 24.85&3.0h&0.0252 & 29.55 & 27.14 & 28.34 & 8.3h&0.0255\\
\hline
11 &  & & - & - & &  & Table \ref{exp: giga_streaming}.1 & 17.52 & 40.49 & 19.91 & 29.27 & 28.66&3.4h&0.0276 & 34.31 & 31.83 & 33.07 & 8.9h&0.0280\\
12 &  & Without  & (A) & 1 $\times$ Uttlen. & Cur. & \multirow{2}{*}{$\leq$20} & Table \ref{exp: giga_streaming}.2 & 17.83 & 40.92 & 20.28 & 29.47 & 29.00&3.6h&0.0310 & 34.50 & 32.13 & 33.32 & 10.0h&0.0313\\
13 &  & utterance & (B) & 3 $\times$ Uttlen. & Uttlen. & \multirow{2}{*}{frames}  & Table \ref{exp: giga_streaming}.5 & 17.12 & $39.86^{\triangle}$ & 18.24 & 28.97 & $28.14^{\triangle}$& 4.5h&0.0364& 33.72 & 31.44 & 32.58 & 11.4h&0.0373\\
14 & \multirow{2}{*}{Streaming} & splicing & (C) & 3 $\times$ 32 &  &  & Table \ref{exp: giga_streaming}.8 &17.27 & 40.02 & 18.43 & 29.16 &28.30&3.6h&0.0312& 33.99 & 31.68 & 32.84 & 10.2h&0.0319\\
\cline{5-6}
15 &  & & (D) & \multicolumn{2}{c|}{$\leq$200 frames}  &  & Table \ref{exp: giga_streaming}.16 & $\textbf{16.82}^{\triangle}$ & $\textbf{39.43}^{\triangle}$ & $\textbf{17.97}$ & $\textbf{28.84}^{\triangle}$ & $\textbf{27.82}^{\triangle}$&3.8h&0.0320 & $\textbf{33.50}^{\triangle}$ & $\textbf{31.15}^{\triangle}$ & $\textbf{32.33}^{\triangle}$ & 10.4h&0.0323\\
\cline{3-20}
16 &  &  & - & - &  &  & Table \ref{exp: giga_streaming}.17 & 17.51 & 40.48 & 19.80 & 29.25 & 28.65 &3.1h& 0.0255& 34.33 & 31.84 & 33.09 & 8.1h&0.0257\\
17 &  & With & (A) & 1 $\times$ Uttlen. &Cur.  & \multirow{2}{*}{$\leq$20} & Table \ref{exp: giga_streaming}.18 & 17.82 & 40.91 & 20.17 & 29.45 & 28.99& 3.3h &0.0265& 34.51 & 32.15 & 33.33 & 9.1h&0.0270\\
18 &  & utterance  & (B) & 3 $\times$ Uttlen. & Uttlen. & \multirow{2}{*}{frames} & Table \ref{exp: giga_streaming}.21 & 17.11 & $39.85^{\star}$ & 18.13 & 28.95& $28.13^{\star}$& 4.1h&0.0325& 33.71 & 31.42 & 32.57 & 10.4h&0.0328\\
19 &  & splicing & (C) & 3 $\times$ 32&  &  & Table \ref{exp: giga_streaming}.24 & 17.26 & 40.01 & 18.32 & 29.14 & 28.29&3.3h&0.0267 & 34.01 & 31.69 & 32.85 & 9.3h&0.0271\\
\cline{5-6}
20 &  &  & (D) & \multicolumn{2}{c|}{$\leq$200 frames}  &  & Table \ref{exp: giga_streaming}.32 & $\textbf{16.81}^{\star}$ & $\textbf{39.42}^{\star}$ &$\textbf{17.86}$ & $\textbf{28.82}^{\star}$ & $\textbf{27.81}^{\star}$&3.5h &0.0285& $\textbf{33.51}^{\star}$ & $\textbf{31.15}^{\star}$ &$\textbf{32.33}^{\star}$ & 9.4h&0.0288\\

\hline\hline
\end{tabular}}
\label{exp: finetuneing_dbank_jcc}
\vspace{-0.3cm}
\end{table*}


\vspace{-0.4cm}
\subsection{Conformer-Transducer Pre-training and Fine-tuning}
\vspace{-0.1cm}
\subsubsection{\textbf{Conformer-Transducer Pre-training Configurations}}
\label{experimental_setup_c_t}
The \textbf{non-streaming baseline} Conformer-Transducer (C-T) systems described in Section~\ref{conformer_transducer} are implemented via Fairseq \cite{ott2019fairseq}. 
80-dimensional mel filterbank (Fbank) features are extracted with global-level cepstral mean and variance normalization. The convolutional subsampling module consists of two 2D convolutional layers with a stride of size 2, each followed by a ReLU activation. The C-T Encoder comprises 12 Encoder blocks, each configured with 8-head attention of 512 dimensions and 2048 feed-forward hidden nodes. The C-T Predictor contains one uni-directional LSTM layer with 300 hidden nodes, while 5000 byte-pair-encoding (BPE) tokens serve as the joint network outputs. SpecAugment \cite{park2019specaugment} is applied for data augmentation. The Adam optimizer is employed with an initial learning rate of 0.0002, 20,000 warmup steps, and a weight decay of 1e-5. The model is trained for 40 epochs, while the final model is obtained by averaging the parameters from the last 5 epochs. During evaluation, both the beam and batch sizes are set to 1, where word or character error rate (WER/CER) are reported for both development and test sets.

The \textbf{streaming baseline C-T} configuration modifies the C-T Encoder by introducing causal convolution in CNN downsampling and chunk-wise self-attention with a look-ahead chunk in each Encoder block. All other components are the same as the non-streaming baseline C-T configuration. The baseline systems operate with a look-ahead chunk of 20 frames (200 ms) within the current utterance. Various chunk sizes are further evaluated for streaming systems, including 60 frames (600 ms), 100 frames (1000 ms), 160 frames (1600 ms), 200 frames (2000 ms), and full utterance processing with complete left-context visibility (denoted as “Cur. Uttlen").

The non-streaming and streaming C-T systems with input audio concatenation further modify the input features as described in Section \ref{input_audio}. Similarly, the non-streaming and streaming C-T systems with cross-utterance Encoder embedding concatenation, cross-utterance Encoder embedding pooling projection, or chunk-based cross-utterance Encoder embeddings apply additional modifications to the Encoder attention module following Section \ref{cross_encoder_emb} to \ref{chunk_based}. All other configurations are the same as the above baseline setup. 

\vspace{-0.03cm}

\subsubsection{\textbf{Conformer-Transducer Fine-tuning on Elderly Speech}}

\label{experiments_elderly_setup}
All the pre-trained C-T systems serving as the starting point for fine-tuning on elderly speech follow the same setup described in Section \ref{experimental_setup_c_t}.
During the fine-tuning stage, \textbf{1)} the pre-trained parameters of the C-T Encoder, Predictor, and Joint network are inherited and fine-tuned on the elderly speech dataset; \textbf{2)} a new linear projection layer is added after the Joint network to reduce the 5000-dimensional output to a 100-dimensional representation for elderly speech. For the English datasets, 5000 BPE tokens are derived from the GigaSpeech M-size dataset, and 100 BPE tokens are extracted from the DementiaBank Pitt elderly speech transcripts. Similar procedures are applied to the Mandarin WenetSpeech M-size pre-training dataset and Cantonese JCCOCC MoCA fine-tuning elderly speech corpus, except that Chinese character-level tokenization is applied instead of BPE tokens. Compared to pre-training, we set a learning rate as 1r-5 with 2,000 warmup steps and a weight decay of 1e-6 for the Adam optimizer. The model is trained over 30 epochs. The rest of configurations are the same as those outlined in Section \ref{experimental_setup_c_t}.

All the experiments are conducted using 8$\times$NVIDIA V100 GPUs. The real-time factor (RTF) is measured on a single V100 GPU for the test sets. 
Statistical significance is assessed using the standard NIST implemented \cite{pallet1990tools} Matched Pairs Sentence-Segment Word Error (MAPSSWE) Test proposed by Gillick \cite{gillick1989some}, with a significant level of $\alpha=0.05$\footnote{\textcolor{black}{Significance tests \tim{are} performed for all WER/CER results. Only those that are significant (MAPSSWE\cite{gillick1989some, pallet1990tools}, $\alpha=0.05$) are highlighted using markers.}}. 



\vspace{-0.3cm}
\section{Experimental Results}
\vspace{-0.1cm}
\label{experiments_results}
In this section, the performance of the four cross-utterance speech contexts modeling approaches are examined for both non-streaming and streaming Conformer-Transducer systems. Experiments are conducted on four benchmark speech tasks across three languages, including: {\bf 1)} the 1000-hour English GigaSpeech \cite{chen2021gigaspeech} M-size and the 1000-hour Mandarin Chinese WenetSpeech \cite{zhang2022wenetspeech} M-size speech corpora for contextual C-T models pre-training; and {\bf 2)} the English DementiaBank Pitt \cite{becker1994natural} and Cantonese JCCOCC MoCA \cite{xu2021speaker} elderly speech datasets for their domain fine-tuning.


Section \ref{performance_cross_nonstreaming} and \ref{performance_cross_streaming} analyze the performance improvements brought by incorporating cross-utterance speech contexts into non-streaming and streaming C-T models during pre-training. The best-performing C-T systems with cross-utterance speech contexts are further compared against 
SOTA systems on the GigaSpeech and WenetSpeech tasks in Section \ref{performance_benchmark_pre-training}. Section \ref{finetuned_wer_performance} to \ref{finetuned_rtf} present the performance of these C-T systems fine-tuned for elderly speech recognition tasks. The best-performing fine-tuned C-T system with cross-utterance speech contexts is further compared against SOTA elderly speech recognition systems in Section~\ref{performance_benchmark_fine-tuning}.
\footnotetext{
We set the dimension of the cross-utterance Encoder embedding pooling projection (Equation~\ref{eqn:pooling}, Section \ref{pooling_projection}) as $L=32$, 
following the best setting in ~\cite{cui23_interspeech} (Table 2, Sys.13 and 16), which offers a better WER-RTF trade-off when compared with other settings, e.g., $L=8$ or $L=16$.}
\vspace{-0.3cm}
\subsection{Performance of Contextual C-Ts on Pre-training Datasets}
\vspace{-0.1cm}

\subsubsection{\textbf{Non-streaming C-T System Results}}
\label{performance_cross_nonstreaming}
In this part, we systematically investigate the performance improvements attributed to the cross-utterance speech contexts fusion methods of Section \ref{context_representation} for non-streaming C-T systems. 
As shown in Table \ref{exp: giga_nonstreaming}, several trends can be observed:


\textbf{i)} The proposed cross-utterance speech contexts fusion approaches (cross-utterance Encoder embedding concatenation (Sys.3-5 and Sys.15-17, fusion method (B) in Table \ref{exp: giga_nonstreaming}), cross-utterance Encoder embedding pooling projection (Sys.6-8 and Sys.18-20, fusion method (C) in Table \ref{exp: giga_nonstreaming}), and chunk-based cross-utterance Encoder embeddings (Sys.9-12 and Sys.21-24, fusion method (D) in Table \ref{exp: giga_nonstreaming})) consistently outperform the baseline C-T systems utilizing the current utterance context only (Sys.1,13) on the GigaSpeech and WenetSpeech pre-training datasets. 
In particular, statistically significant WER reductions of \textbf{0.9\%}, \textbf{0.9\%}, \textbf{0.9\%} absolute (\textbf{6.0\%}, \textbf{6.0\%}, \textbf{6.0\%} relative) are obtained by the C-T systems with cross-utterance Encoder embedding concatenation over the corresponding baseline (Sys.5 vs. Sys.1) across GigaSpeech \textbf{DEV}, \textbf{TEST}, and \textbf{Avg.} sets, while CER reduction of \textbf{1.1\%} absolute (\textbf{5.4\%} relative) is obtained on WenetSpeech \textbf{TEST} set (Sys.5 vs. 1).

\textbf{ii)} When comparing the cross-utterance speech contexts fusion approaches, cross-utterance Encoder embedding concatenation (Sys.3-5 and Sys.15-17, fusion method (B) in Table \ref{exp: giga_nonstreaming}) achieves better performance than cross-utterance Encoder embedding pooling projection (Sys.6-8 and Sys.18-20, fusion method (C) in Table \ref{exp: giga_nonstreaming}) and chunk-based cross-utterance Encoder embedding (Sys.9-12 and Sys.21-24, fusion method (D) in Table \ref{exp: giga_nonstreaming}). 
This can be attributed to the modeling consistency between previous and current utterance contexts when utilizing the cross-utterance speech contexts approaches.
In contrast, the cross-utterance Encoder embedding pooling projection and chunk-based cross-utterance Encoder embedding approaches produce partial, compressed, or truncated cross-utterance speech contexts.

\textbf{iii)} The proposed efficient batch-mode training with utterance splicing improves the efficiency across all contextual or non-contextual C-T systems. This improvement is achieved by splicing multiple neighbouring utterances within minibatches to minimize the synchronization overhead during batch-mode training.
Experiments demonstrate that our efficient batch-mode training with utterance splicing reduces training time by \textbf{9.8\%} to \textbf{16.6\%} (Sys.24 vs. Sys.12 and Sys.18 vs. Sys.6 on the GigaSpeech dataset in Table \ref{exp: giga_nonstreaming}), and \textbf{7.5\%} to \textbf{15.5\%} (Sys.17 vs. Sys.5 and Sys.13 vs. Sys.1 on the WenetSpeech dataset in Table \ref{exp: giga_nonstreaming}), 
highlighting a large improvement in training efficiency. These results demonstrate that batch-mode training with utterance splicing effectively increases GPU memory utilization while preserving cross-utterance speech contexts in a left-to-right monotonic manner. 
Such a method can also be applied to non-C-T based ASR systems to improve the training efficiency when modeling cross-utterance speech contexts.

\textbf{iv)} Increasing 
the number of previous utterances (from 1 to 2 or 3) or the number of frames in previous utterances (from 60 to 100, 160, or 200) consistently yields improvements over modeling only the most recent utterance (Sys.4,5 vs. Sys.3; Sys.7,8 vs. Sys.6; Sys.10-12 vs. Sys.9; Sys.16,17 vs. Sys.15; Sys.19,20 vs. Sys.18; Sys.22-24 vs. Sys.21 in Table \ref{exp: giga_nonstreaming}). Such gains are consistent across all three cross-utterance speech contexts modeling approaches (B), (C), and (D) in Table \ref{exp: giga_nonstreaming}.

\textbf{v)} The three cross-utterance speech contexts fusion approaches (cross-utterance Encoder embedding concatenation (Sys.3-5 and Sys.15-17, (B) in Table \ref{exp: giga_nonstreaming}), cross-utterance Encoder embedding pooling projection (Sys.6-8 and Sys.18-20, (C) in Table \ref{exp: giga_nonstreaming}), and chunk-based cross-utterance Encoder embeddings (Sys.9-12 and Sys.21-24, (D) in Table \ref{exp: giga_nonstreaming})) introduce minimal impact on computational efficiency. Such marginal increases in computation overhead demonstrate that the three context fusion approaches can achieve WER/CER reductions without notably impacting the efficiency of systems. Among these, the cross-utterance Encoder embedding pooling projection method incurs the smallest increase in computation (measured in RTFs) by \textbf{3.3\%} and \textbf{2.1\%} over the baseline C-T system respectively on the GigaSpeech and WenetSpeech datasets (Sys.6 vs Sys.1 in Table \ref{exp: giga_nonstreaming}).


\begin{table*}[htbp]
\caption{Performance (WER/CER\%) contrasts between published systems and contextual C-T systems using cross-utterance speech contexts 
on the ELDERLY English DEMENTIABANK (DB.) and the Cantonese JCCOCC MOCA (JCC.) Corpora. Here (B) denotes cross-utterance Encoder embedding \tim{concatenation} of Section \ref{cross_encoder_emb}, while (D) is chunk-based cross-utterance Encoder embeddings of Section \ref{chunk_based}. “LM" represents language model,  “MLAN" corresponds to multi-level adaptive networks, and “VR-SBE" means variance regularized spectral basis embedding. “LS.", “GS-M", “WS-M" and “LL." respectively stand for the LibriSpeech, the GigaSpeech M-size, the WenetSpeech M-size, and the LibriLight datasets. “Params" represents parameters.}
\vspace{-0.3cm}

\centering
\resizebox{\linewidth}{!}{
\setlength{\tabcolsep}{2pt} 
\fontsize{7.5}{10}
\begin{tabular}{c|c|c|c|c|c|c|c|c|cc|c|c|c|c} 
\hline\hline
\multirow{3}{*}{ID} & 
\multirow{3}{*}{System} &
\multicolumn{2}{c|}{Training data }
 &
 \multirow{3}{*}{\#Params.} &
  \multirow{3}{*}{\textcolor{black}{\#FLOPs}} &
\multirow{2}{*}{Fusion} &
\multicolumn{5}{c|}{Dementiabank Pitt WER} &
\multicolumn{3}{c}{JCCOCC MoCA CER }  \\
\cline{3-4}
\cline{8-15}
& & \multirow{2}{*}{Source} & \multirow{2}{*}{Size (\#hour)}&&&\multirow{2}{*}{method} &\multicolumn{2}{c|}{DEV} &\multicolumn{2}{c|}{EVAL} & \multirow{2}{*}{Avg.}& \multirow{2}{*}{DEV} & \multirow{2}{*}{EVAL} & \multirow{2}{*}{Avg.}  \\

\cline{8-11}
&   & & &&&&    INV.& PAR. & INV. &PAR. &  & & &  \\
\hline\hline

1 & 2021 CUHK Kaldi TDNN \cite{ye2021development}  & DB. & 0.06k / 0.16k&18M&\textcolor{black}{-}&-&19.91&47.93&19.76&36.66&33.80 & \multicolumn{3}{c}{-}\\

2 & 2021 CUHK Kaldi TDNN + Bayesian Adaptation + 4-gram LM \cite{ye2021development}& DB. & 0.06k &18M&\textcolor{black}{-}&-&17.61 & 42.12 & 17.20 & 33.17 & 29.90& \multicolumn{3}{c}{-}\\
3 & 2022 CUHK Conformer\cite{hu2022exploitinguti} &DB. / JCC. & 0.06k / 0.16k& 53M&\textcolor{black}{$3.730 \times10^{11}$}&-& 20.97&48.71& 19.42&36.93& 34.57& 33.08& 31.24 & 32.15 \\
4 & 2022 CUHK Conformer + MLAN + Score Fusion \cite{hu2022exploitinguti} &DB. / JCC. & 0.06k / 0.16k&53M&\textcolor{black}{-}&-& 20.38&47.68& 17.87&36.13& 33.75 & 31.78 & 29.93 & 30.85\\
5 & 2022 CUHK Conformer + VR-SBE Adaptation\cite{geng2025homogeneous} &DB. / JCC. & 0.06k & 53M&\textcolor{black}{-}&-& 20.83&47.39& 17.64&36.34& 33.84 & 32.42 & 31.01 & 31.71\\
6 & 2024 CUHK Whisper-Medium &Mixed data & 680k&769M&\textcolor{black}{$1.294\times10^{12}$}&-&19.89 &37.92 &17.31 &26.24& 28.01 & \multicolumn{3}{c}{-}\\
7 & 2024 CUHK Whisper-Medium + Lora Fine-tuning  &Mixed data + DB. / JCC.& 680k + 0.06k / 0.16k&769M&\textcolor{black}{$1.301\times10^{12}$}&-& 12.76 &28.79 &12.65 & 20.68& 20.43 &28.68 &25.79 &27.23 \\
8 & 2024 CUHK Wav2vec2.0-Conformer + Domain Adaptation \cite{hu2024self}  &LL. + DB.& 60k + 0.06k&619M&\textcolor{black}{$1.182\times10^{12}$}&-& 14.29  &29.71 & 15.32 &  21.27& 21.60& \multicolumn{3}{c}{-}\\
9 & 2024 CUHK XLSR-128 + Domain Adaptation\cite{hu2024self}  &Mixed data + JCC.& 436k + 0.16k&300M&\textcolor{black}{$7.452 \times10^{11}$}&-&  \multicolumn{5}{c|}{-} & 28.86 & 26.53 & 27.69\\
10 & 2021 CUHK Kaldi TDNN + Bayesian Adaptation \cite{deng21d_interspeech}& LS. + DB.& 0.96k + 0.06k &18M&\textcolor{black}{-}&-&19.07 & 43.36 & 17.98 & 32.08 & 30.83& \multicolumn{3}{c}{-}\\
11 & 2022 CUHK Conformer + Domain\&Speaker Adaptation \cite{wang22k_interspeech} & LS. + DB.&  0.96k + 0.06k&53M&\textcolor{black}{-}&-&16.0&35.2   & 15.3& 26.4  & 25.3 & \multicolumn{3}{c}{-}\\
12 & 2023 CUHK Conformer + Hyper-parameter Adaptation \cite{wang2023hyper}  & LS. + DB. &  0.96k + 0.06k&53M&\textcolor{black}{-}&-& 15.20 & 34.71 &14.76&  25.80  & 24.68 & \multicolumn{3}{c}{-}\\
\hline
13 & \textbf{Our non-streaming C-T of Section \ref{cross_encoder_emb} (Sys.8 in Table \ref{exp: finetuneing_dbank_jcc})} & GS-M + DB. / WS-M + JCC.& 1k + 0.06k / 1k + 0.16k&88.5M&\textcolor{black}{$4.97 \times10^{11}$}& (B)&  \textbf{14.99} &  \textbf{34.95} &  \textbf{15.45} &  \textbf{25.49} &  \textbf{24.68} &  \textbf{29.12} &  \textbf{26.96} &  \textbf{28.04} \\
14 & \textbf{Our streaming C-T of Section \ref{chunk_based} (Sys.20 in Table \ref{exp: finetuneing_dbank_jcc})}  & GS-M + DB. / WS-M + JCC. & 1k + 0.06k / 1k + 0.16k&88.5M&\textcolor{black}{$5.23 \times10^{11}$}& (D)&  \textbf{16.81} &  \textbf{39.42} &  \textbf{17.86} &  \textbf{28.82} &  \textbf{27.81} &  \textbf{33.51} &  \textbf{31.15} &  \textbf{32.33} \\

\hline\hline
\end{tabular}}
\label{exp: existing_fine-tuning}
\vspace{-0.3cm}
\end{table*}

\subsubsection{\textbf{Streaming C-T System Results}}
\label{performance_cross_streaming}
When applying cross-utterance speech fusion methods (Section \ref{context_representation}) in streaming C-T systems, several similar trends can be observed in Table \ref{exp: giga_streaming}:


\textbf{i)} The streaming C-T systems with the cross-utterance speech contexts fusion approaches, including cross-utterance Encoder embedding concatenation (Sys.3-5, 19-21, (B) in Table \ref{exp: giga_streaming}), cross-utterance Encoder embedding pooling projection (Sys.6-8, 22-24, (C) in Table \ref{exp: giga_streaming}), and chunk-based cross-utterance Encoder embeddings (Sys.13-16, 29-32, (D) in Table \ref{exp: giga_streaming}) consistently outperform the baseline systems solely relying on the current utterance context (Sys.1, 17 in Table \ref{exp: giga_streaming}). 

\textbf{ii)} 
All the C-T systems with chunk-based cross-utterance Encoder embedding (Sys.13-16, 29-32, (D) in Table \ref{exp: giga_streaming}) outperform the non-contextual streaming C-T systems with statistically significant average WER reductions 
from \textbf{0.5}\% (Sys.29 vs. Sys.25 in Table \ref{exp: giga_streaming}) to \textbf{1.0}\% (Sys.32 vs. Sys.28 in Table \ref{exp: giga_streaming}) absolute (\textbf{2.8}\% to \textbf{5.7}\% relative), and CER reductions 
from \textbf{0.46}\% (Sys.14 vs. Sys.10 in Table \ref{exp: giga_streaming}) to \textbf{0.62}\% (Sys.32 vs. Sys.28 in Table \ref{exp: giga_streaming}) absolute (\textbf{1.9}\% to \textbf{2.6}\% relative) on the GigaSpeech \textbf{Avg.} and WenetSpeech \textbf{TEST} sets, respectively.

\textbf{iii)} Among the three cross-utterance speech modeling approaches, the chunk-based cross-utterance Encoder embedding approach (Sys.13-16, 29-32, fusion method (D) in Table \ref{exp: giga_streaming}) is found to be the most effective for the streaming C-T system, outperforming the other two approaches in terms of WER/CER on the GigaSpeech \textbf{Avg.} and WenetSpeech \textbf{TEST} sets. In particular, the proposed chunk-based cross-utterance Encoder embedding approach (Sys.16 in Table \ref{exp: giga_streaming}) outperforms the baseline C-T system (Sys.1 in Table \ref{exp: giga_streaming}) by statistically significant WER reductions of \textbf{1.0}\%, \textbf{1.0}\%, \textbf{1.0}\% absolute (\textbf{5.6}\%, \textbf{5.7}\%, \textbf{5.7}\% relative), and CER reductions of \textbf{1.0}\% absolute (\textbf{4.2}\% relative) on the GigaSpeech \textbf{DEV}, \textbf{TEST}, \textbf{Avg.} and WenetSpeech \textbf{TEST} sets. In contrast, using either the cross-utterance Encoder embedding concatenation (Sys.3-5, 19-21, fusion method (B) in Table \ref{exp: giga_streaming}) or cross-utterance Encoder embedding pooling projection (Sys.6-8, 22-24, fusion method (C) in Table \ref{exp: giga_streaming})
underperforms the systems with the chunk-based cross-utterance Encoder embedding approach (Sys.13-16, 29-32 in Table \ref{exp: giga_streaming}) in streaming mode. This may be attributed to the modeling inconsistency between previous utterance and current utterance contexts.

\textbf{iv)} Regarding training efficiency, all the C-T models trained in batch mode using within-minibatch utterance splicing achieve faster training speed than those trained without it (Sys.17 vs. Sys.1; Sys.19-21 vs. Sys.3-5; Sys.22-24 vs. Sys.6-8; Sys.29-32 vs. Sys.13-16 in Table \ref{exp: giga_streaming}). 
Consistent model training time reduction of \textbf{12.5\%} to \textbf{18.6\%} (Sys.4 vs. Sys.20 and Sys.13 vs. Sys.29 in Table \ref{exp: giga_streaming}) on the GigaSpeech dataset, and \textbf{8.9\%} to \textbf{17.9\%} (Sys.4 vs. Sys.20 and Sys.1 vs. Sys.17 in Table \ref{exp: giga_streaming}) on the WenetSpeech dataset) are obtained.

\textbf{v)} Further increasing the number of previous utterances or frames from 1 to 2, 3, or from 60 to 100, 160, 200, consistently leads to better performances compared to solely modeling the most recent utterance (Sys.4,5 vs. Sys.3; Sys.7,8 vs. Sys.6; Sys.14-16 vs. Sys.13; Sys.20,21 vs. Sys.19; Sys.23,24 vs. Sys.22; Sys.30-32 vs. Sys.29 in Table \ref{exp: giga_streaming}).

\textbf{vi)} 
Similar to the trends of non-streaming systems presented in \textbf{v)} of Section \ref{performance_cross_nonstreaming}, none of the cross-utterance speech context fusion approaches (B), (C), and (D) in Table \ref{exp: giga_streaming} lead to a substantial increase in C-T inference RTFs during streaming mode.
Among these, the cross-utterance Encoder embedding pooling projection approach 
(Sys.6, fusion method (C) in Table \ref{exp: giga_streaming}) incurs the smallest increase in RTFs during Encoder context fusion by \textbf{2.4}\% and \textbf{3.1\%} over the baseline C-T system respectively on the GigaSpeech and WenetSpeech datasets (Sys.6 vs. Sys.1 in Table \ref{exp: giga_streaming}).


While we have attempted to implement MERL's input audio context feature fusion method~\cite{hori2020transformer, hori2021advanced} (Sys.2, 14 in Table \ref{exp: giga_nonstreaming}; Sys.2, 18 in Table \ref{exp: giga_streaming}) for C-T models,\footnote{Our correspondence with the authors indicates that their implementation is proprietary and not publicly available.} 
our results thus far suggest that this approach does not perform as well as the three cross-utterance contexts fusion methods of our paper.

\subsubsection{\textbf{Performance Contrast against SOTA in Pre-training}}
\label{performance_benchmark_pre-training}
The performance of the best pre-trained systems 
with cross-utterance speech contexts is further compared with 
SOTA performance on the English GigaSpeech M-size and Mandarin WenetSpeech M-size tasks. As shown in Table \ref{exp: existing_pre-training}:

\textbf{i)} On the English GigaSpeech task, our method with cross-utterance Encoder embeddings of Section \ref{cross_encoder_emb}
(Sys.8, fusion method (B) in Table \ref{exp: existing_pre-training})
achieves comparable WER reductions 
to the LongFNT offline system \cite{gong2024advanced, gong2023longfnt} (Sys.3 in Table \ref{exp: existing_pre-training}), while the latter using over \textbf{10} times more pre-training data. 

\textbf{ii)} On the Mandarin WenetSpeech task, 
the proposed cross-utterance Encoder embeddings of Section \ref{cross_encoder_emb} 
(Sys.8, fusion method (B) in Table \ref{exp: existing_pre-training})
achieves a \textbf{14.55\%} absolute (\textbf{42.9\%} relative) CER reduction compared to the Whisper-Medium 
model \cite{radford2023robust} (Sys.2 in Table \ref{exp: existing_pre-training}), while using just \textbf{0.15\%} of its pre-training data and \textbf{88.5\%} fewer model parameters.

\vspace{-0.3cm}
\subsection{Performance of Contextual C-Ts on Elderly Speech Data}
\vspace{-0.1cm}
\label{performance_finetune}


In this section, we utilize the best-performing pre-trained C-T systems (Sys.2, 5, 8, 12, 14, 17, 20, 24 in Table \ref{exp: giga_nonstreaming}, and Sys.2, 5, 8, 16, 18, 21, 24, 32 in Table \ref{exp: giga_streaming}) as starting points for fine-tuning on elderly speech. Experiments are conducted on two benchmark elderly speech recognition tasks, including the English DementiaBank and Cantonese JCCOCC MoCA elderly speech datasets. Trends similar to those observed on the pre-training datasets can be found on three fronts:
\begin{figure}
    \centering
    \includegraphics[width=0.9\linewidth]{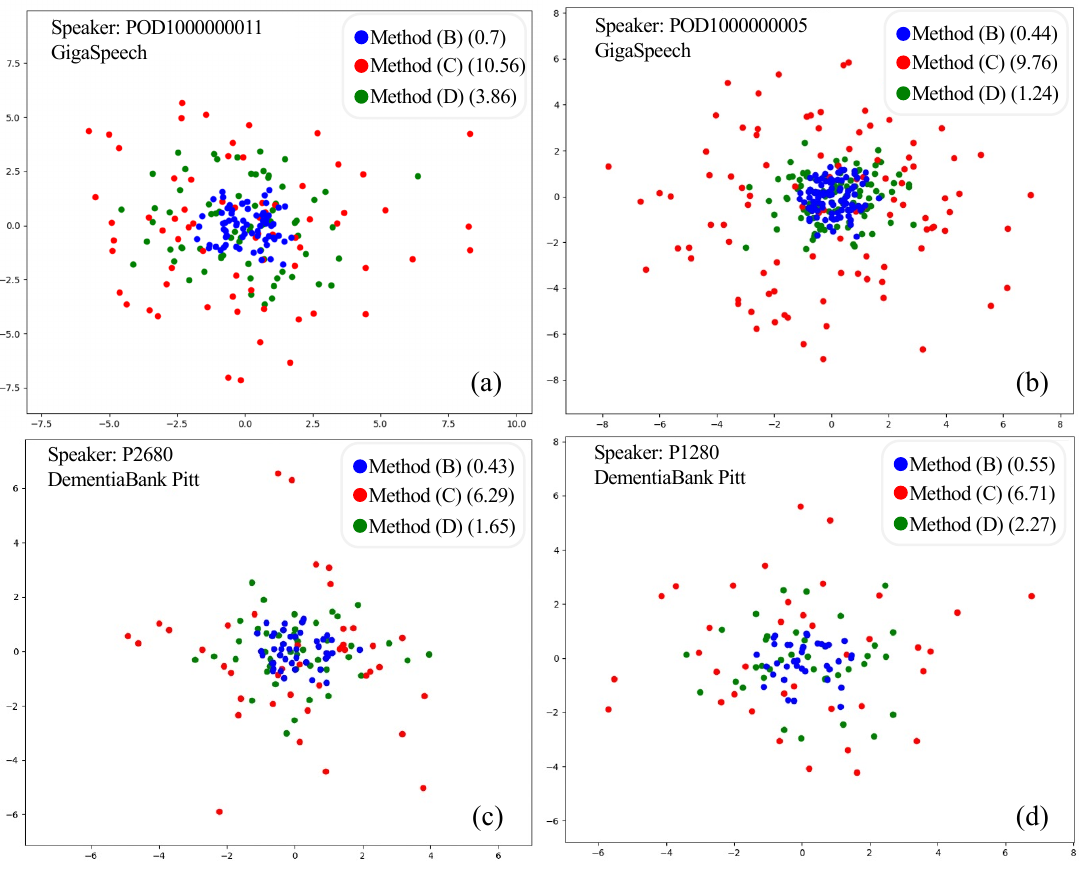}
    \caption{\textcolor{black}{T-SNE visualizations of the context-fused encoder representations obtained using the three proposed fusion methods after the 12th Encoder transformer block: cross-utterance Encoder embedding concatenation (method (B)), cross-utterance Encoder embedding pooling projection (method (C)) and chunk-based cross-utterance Encoder embeddings (method (D)). Speakers POD1000000011 and POD1000000005 are from the GigaSpeech corpus, while P2680 and P1280 are from the DementiaBank Pitt elderly speech dataset. The determinant of the covariance matrix for each fusion method derived cross-utterance speech contexts representations is indicated in brackets.} }
    \label{fig:tsne}
\end{figure}
\subsubsection{\textbf{WER/CER Performance}} 
\label{finetuned_wer_performance}
The best pre-trained contextual C-T systems (i.e., cross-utterance Encoder embedding concatenation (Sys.5, 17 
in Table \ref{exp: giga_nonstreaming} and Sys.5, 21 
in Table \ref{exp: giga_streaming}), cross-utterance Encoder embedding pooling projection (Sys.8, 20 
in Table \ref{exp: giga_nonstreaming} and Sys.8, 24 
in Table \ref{exp: giga_streaming}), and chunk-based cross-utterance Encoder embeddings (Sys.12, 24 
in Table \ref{exp: giga_nonstreaming} and Sys.16, 32 
in Table \ref{exp: giga_streaming})) consistently outperform non-contextual C-T baselines after fine-tuning on the elderly speech datasets.
This consistency is observed over a range of C-T systems with cross-utterance speech contexts fusion approaches (B), (C), and (D): \textbf{i)} non-streaming C-T systems trained without within-minibatch utterance splicing (Sys.3-5 in Table \ref{exp: finetuneing_dbank_jcc}); \textbf{ii)} non-streaming C-T systems with utterance splicing (Sys.8-10 in Table \ref{exp: finetuneing_dbank_jcc}); \textbf{iii)} streaming C-T systems without utterance splicing (Sys.13-15 in Table \ref{exp: finetuneing_dbank_jcc}); and \textbf{iv)} streaming C-T systems with utterance splicing (Sys.18-20 in Table \ref{exp: finetuneing_dbank_jcc}). These systems consistently outperform their corresponding non-contextual baselines (Sys.1, 6, 11, 16 in Table \ref{exp: finetuneing_dbank_jcc}). In particular, statistically significant average WER reductions of \textbf{0.51\%} absolute (\textbf{2.0\%} relative),
and average CER reductions of \textbf{0.98\%} absolute (\textbf{3.4\%} relative) are respectively obtained by the non-streaming C-T systems with cross-utterance Encoder embedding concatenation over the non-contextual baselines (Sys.8 vs. Sys.6 in Table \ref{exp: finetuneing_dbank_jcc}). 
\textcolor{black}{As shown in Figure~\ref{fig:tsne}, these WER reductions align with the more invariant context-fused Encoder representations produced by cross-utterance Encoder embedding concatenation (method (B), in blue) in the t-SNE visualizations, compared to those generated by cross-utterance Encoder embedding pooling projection (method (C), in green) and chunk-based cross-utterance Encoder embeddings (method (D), in red)}.

\subsubsection{\textbf{Model Fine-tuning Efficiency}} 
\label{finetuned_efficiency}

The proposed efficient batch-mode training scheme consistently speeds up C-T model fine-tuning on the elderly speech datasets. 
For non-streaming systems, The maximum training time reduction is \textbf{15.2\%} (Sys.9 vs. Sys.4 in Table \ref{exp: finetuneing_dbank_jcc}) on the DementiaBank Pitt dataset and \textbf{9.9\%} (Sys.8 vs. Sys.3 in Table \ref{exp: finetuneing_dbank_jcc}) on the JCCOCC MoCA corpus. For streaming systems, the maximum reduction is \textbf{8.9\%} (Sys.18 vs. Sys.13 in Table \ref{exp: finetuneing_dbank_jcc}) on DementiaBank Pitt and \textbf{9.6\%} (Sys.20 vs. Sys.15 in Table \ref{exp: finetuneing_dbank_jcc}) on JCCOCC MoCA.


\subsubsection{\textbf{Inference RTF}} 
\label{finetuned_rtf}

For both non-streaming and streaming systems, none of the cross-utterance speech contexts fusion approaches (B), (C), and (D) in Table \ref{exp: finetuneing_dbank_jcc} lead to a substantial increase in C-T inference RTFs after their fine-tuning on the elderly speech datasets (Sys.2-5 vs. Sys.1; Sys.7-10 vs. Sys.6; Sys.12-15 vs. Sys.11; Sys.17-20 vs. Sys.16 in Table \ref{exp: finetuneing_dbank_jcc}).
The lowest RTF is obtained by the non-streaming C-T system with cross-utterance Encoder embedding pooling projection
(Sys.9, fusion method (C) in Table \ref{exp: finetuneing_dbank_jcc}), incurring a small increase of \textbf{4.2\%} in RTF on Dementiabank (\textbf{3.3\%} on JCCOCC MoCA) over the baseline (Sys.9 vs. Sys.6, Table \ref{exp: finetuneing_dbank_jcc}). \textcolor{black}{
\textcolor{black}{In addition, as shown in Table~\ref{exp: existing_fine-tuning}, compared to large foundation models (Sys.6-9), our proposed systems (Sys.13,14) achieve comparable WER/CER performance with far fewer parameters and lower floating point operations (FLOPs), highlighting their practicality for resource-constrained deployment.} }


\subsubsection{\textbf{Performance Contrast against SOTA after Fine-tuning}}
\label{performance_benchmark_fine-tuning}
The performance of our best elderly speech fine-tuned contextual C-T systems 
are compared against SOTA systems for elderly speech recognition. Table~\ref{exp: existing_fine-tuning} shows several trends:

\textbf{i)} Our non-streaming contextual C-T system with cross-utterance Encoder embeddings of Section \ref{cross_encoder_emb}
(Sys.13, fusion method (B) in Table \ref{exp: existing_fine-tuning}) achieves comparable WER performance to existing systems on 
Dementiabank Pitt when using a similar amount of training data (Sys.10-12 in Table \ref{exp: existing_fine-tuning}).

\textbf{ii)} Our non-streaming contextual C-T system with cross-utterance Encoder embedding concatenation (Sys.13 in Table \ref{exp: existing_fine-tuning}) achieves comparable WER/CER performance on both elderly datasets compared to the XLSR-128 (Sys.9 in Table \ref{exp: existing_fine-tuning}) and Whisper-Medium (Sys.6,7 in Table \ref{exp: existing_fine-tuning}) foundation models (\textbf{$<$0.87\%} relative CER increase), while using less than \textbf{0.15\%} of their training data and \textbf{88.5\%} fewer model parameters (Sys.13 vs. Sys.7 in Table \ref{exp: existing_fine-tuning}).

\textbf{iii)} Compared to the Whisper-Medium system fine-tuned with LoRA (Sys.2 in Table 
\ref{exp: existing_pre-training} and Sys.7 in Table \ref{exp: existing_fine-tuning}), our 
contextual C-T system, which incorporates cross-utterance Encoder embeddings of Section \ref{cross_encoder_emb} 
(Sys.8, fusion method (B) in Table 
\ref{exp: existing_pre-training} and Sys.13 in Table \ref{exp: existing_fine-tuning}), 
leads to larger WER reductions and comparable CER performance on the Chinese Mandarin WenetSpeech and the Cantonese JCCOCC MoCA elderly speech datasets. These results highlight the potential of integrating longer-range, cross-utterance speech contexts into current speech foundation models.



\vspace{-0.3cm}
\section{Conclusion}
\vspace{-0.1cm}
\label{conclusion}

In this paper, we present the first extensive study of a range of cross-utterance speech contexts modeling approaches for streaming and non-streaming Conformer-Transducer (C-T) models that are based on either: \textbf{i}) input audio feature concatenation; \textbf{ii}) cross-utterance Encoder embedding concatenation; \textbf{iii}) cross-utterance Encoder embedding pooling projection; or \textbf{iv}) novel chunk-based cross-utterance Encoder embeddings.
To minimize the synchronization overhead during parallel training, an efficient batch-training scheme using contiguous blocks of spliced utterances within each minibatch is further proposed.
Experiments on four benchmark speech datasets across three languages (the English GigaSpeech and Mandarin Wenetspeech corpora used in contextual C-T models pre-training; and the English DementiaBank Pitt and Cantonese JCCOCC MoCA elderly speech datasets used in domain fine-tuning) suggest that incorporating cross-utterance speech contexts improves the ASR system's performance on both normal and low-resource elderly speech.
The best-performing streaming C-T models with cross-utterance speech contexts outperform their respective baselines by statistically significant average WER or CER reductions to \textbf{0.9\%}, \textbf{1.1\%}, \textbf{0.51\%}, and \textbf{0.98\%} absolute (\textbf{6.0\%}, \textbf{5.4\%}, \textbf{2.0\%}, and \textbf{3.4\%} relative) on the four tasks.
Their competitive performance against Wav2vec2.0-Conformer, XLSR-128, and Whisper models highlights the potential of integrating cross-utterance speech contexts into current speech foundation models. 
Future work will focus on cross-utterance speech contexts modeling for LLM-based ASR and pathological speech analysis.
\vspace{-0.3cm}

\bibliographystyle{IEEEtran}
\bibliography{IEEEexample}

\end{document}